\documentclass[reprint,amsmath,amssymb,aps,superscriptaddress]{revtex4-2}
\usepackage{graphicx}
\usepackage{dcolumn}
\usepackage{bm}
\usepackage{dsfont}
\usepackage{amsmath}
\usepackage{braket}
\usepackage{courier}
\usepackage{graphicx}
\usepackage{cancel}
\usepackage[caption=false]{subfig}
\usepackage{color}
\usepackage[normalem]{ulem}
\usepackage[hidelinks]{hyperref}
\usepackage{float}

\newcommand{\OPc}[2]{\hat{#1}_{#2}^{\dag}}
\newcommand{\OP}[2]{\hat{#1}_{#2}^{\vphantom{\dag}}}
\newcommand{\CD}[1]{\OPc{c}{#1}}
\newcommand{\C}[1]{\OP{c}{#1}}
\newcommand{\ND}[1]{\hat{n}_{#1}}

\newcommand{\conditionalGraphics}[1]{#1}

\begin{document}

\title{Light-Induced Control of Magnetic Phases in Kitaev Quantum Magnets}%


%

\author{Adithya Sriram}
\affiliation{Department of Physics and Astronomy, University of Pennsylvania, Philadelphia, PA 19104}
\affiliation{Center for Computational Quantum Physics, Flatiron Institute, Simons Foundation, New York City, NY 10010}
\author{Martin Claassen}%
 \email{claassen@sas.upenn.edu}
\affiliation{Department of Physics and Astronomy, University of Pennsylvania, Philadelphia, PA 19104}
\affiliation{Center for Computational Quantum Physics, Flatiron Institute, Simons Foundation, New York City, NY 10010}

\newcommand{\captiontitle}[1]{\textbf{#1}}

\newcommand{\mysection}[1]{\paragraph*{#1 ---}}

\date{\today}

\begin{abstract}
	Leveraging coherent light-matter interaction in solids is a promising new direction towards control and functionalization of quantum materials, to potentially realize regimes inaccessible in equilibrium and stabilize new or useful states of matter. We show how driving the strongly spin-orbit coupled proximal Kitaev magnet $\alpha$-RuCl$_3$ with circularly-polarized light can give rise to a novel ligand-mediated magneto-electric effect that both photo-induces an effective magnetic field and dramatically alters the interplay of competing exchange interactions. We propose that tailored light pulses can nudge the material towards the elusive Kitaev quantum spin liquid as well as probe competing magnetic instabilities far from equilibrium, and predict that the transient competition of magnetic exchange processes can be readily observed via pump-probe spectroscopy.
\end{abstract}

\maketitle

The search for elusive correlated and topological phases of matter, such as the Kitaev quantum spin liquid (QSL), is driven by fundamental insights and tantalizing applications that derive from their exotic properties \cite{keimer17}. However, their realization typically requires fine control over lattice, electronic and magnetic interactions, a fundamental challenge in materials discovery and design \cite{basov17}. 

An intriguing possibility involves harnessing the coherent interaction between light and matter using tailored laser pulses to drive a material out of equilibrium and modify its properties \cite{basov17,delatorre21,oka18}. Seminal time-resolved experiments demonstrated coherent ``dressing'' of electronic bands in solids for the duration of a pump pulse \cite{wang13,mahmood16,mciver20}, suggesting that such ``Floquet engineering'' could be fruitfully extended to strongly-correlated systems \cite{bukov15b,mentink15,claassen17}. Here, light can serve to selectively modify the interplay of competing interactions and realize states of matter not readily accessible in equilibrium.

Beyond Mott-Hubbard physics, recent generalizations to multi-orbital systems investigated light-induced spin-orbital dynamics \cite{liu18,hejazi18,chaudhary19} or tunable exchange anisotropy and Kitaev interactions in the presence of spin-orbit coupling \cite{arakawa21}. However, a central challenge remains to marry Floquet engineering with the complexity of real transition metal compounds, as exemplified by recent work \cite{PhysRevLett.125.197203,chaudhary2020controlling} on the crucial role of ligands in optically controlling magnetic states. At the same time, this added materials complexity could potentially open new pathways to selectively address electronic or magnetic properties.

Here, we predict that circularly-polarized optical irradiation of a strongly spin-orbit coupled (SOC) magnet, $\alpha$-RuCl$_3$ -- a candidate Kitaev material -- grants a novel two-fold handle to control the magnetic state. First, the chirally-oscillating electric field component can induce a large ligand-mediated effective magnetic field in the [111] direction, the field direction required to induce a gap in the Kitaev model. Second, an optical pulse can selectively modify the interplay between ligand superexchange and direct exchange, thereby dramatially altering the balance between isotropic and anisotropic magnetic interactions. These effects can conspire to drive the material towards a non-equilibrium realization of the Kitaev QSL. We predict that resulting transient modifications of the magnetic dynamics can be readily observed in time-resolved probes of the excitation spectrum, and find a host of competing instabilities that can be probed out of equilibrium.

\mysection{Light-matter interaction in Kitaev magnets} In Kitaev materials such as RuCl\textsubscript{3}, a single hole resides in the $t_{2g}$ manifold of the transition metal (TM) $d$ orbitals, which splits into $j_{\textrm{eff}} = 1/2$ and $j_{\textrm{eff}} = 3/2$ states via strong SOC to form an effective spin-$1/2$ local moment \cite{jackeli09,chaloupka12,rau14,PhysRevB.93.214431,Winter_2017,PhysRevB.95.024426,PhysRevB.91.144420}. Fortuitous interference of ligand-mediated superexchange processes along the ~$\sim 90^\circ$ Ru-Cl-Ru bond gives rise to the Kitaev interaction $K \hat{S}_i^\gamma \hat{S}_j^\gamma$, with $\gamma = x,y,z$ along the three inequivalent bonds that form a honeycomb lattice. Conversely, subdominant Heisenberg and off-diagonal anisotropic interactions arise mainly from direct TM-TM exchange and instead induce conventional zigzag magnetic order in RuCl\textsubscript{3}.

The geometric dependence of exchange processes underlying each of the competing magnetic interactions, combined with strong SOC, suggests that tailored light pulses can in principle selectively enhance or suppress certain desired processes and nudge the material towards a QSL. To describe this effect, consider $\alpha-$RuCl\textsubscript{3} irradiated with a broad circularly-polarized light pulse, with the plane of polarization in the octahedral plane. For a single octahedron, the dynamics of holes in the Ru $t_{2g}$ orbitals and charge-transfer (CT) excitations to the edge-sharing ligand $p$ orbitals are given by \cite{PhysRevB.93.214431}:
\begin{align}
    \hat{H}_0 &= U \sum_{i\alpha} \ND{i\alpha\uparrow} \ND{i\alpha\downarrow} + \sum_{i\sigma\sigma',\alpha<\beta} (U' - \delta_{\sigma\sigma'} J_H) \ND{i\alpha\sigma'} \ND{i\beta\sigma} \notag\\
    &+ J_H \sum_{i,\alpha \neq \beta} \left( \CD{i\alpha\uparrow} \CD{i\alpha\downarrow} \C{i\beta\downarrow} \C{i\beta\uparrow} - \CD{i\alpha\uparrow} \C{i\alpha\downarrow} \CD{i\beta\downarrow} \C{i\beta\uparrow} \right) \notag\\
    &+ \frac{\lambda}{2} \sum_i \mathbf{c}^\dag_i (\mathbf{L}\cdot \mathbf{S}) \mathbf{c}_i ~+~ \Delta \sum_{i'} \OPc{p}{i'\sigma} \OP{p}{i'\sigma}
\end{align}
\begin{figure}[h]
    \conditionalGraphics{\includegraphics[width=\columnwidth,trim=1cm 0cm 0cm 0cm]{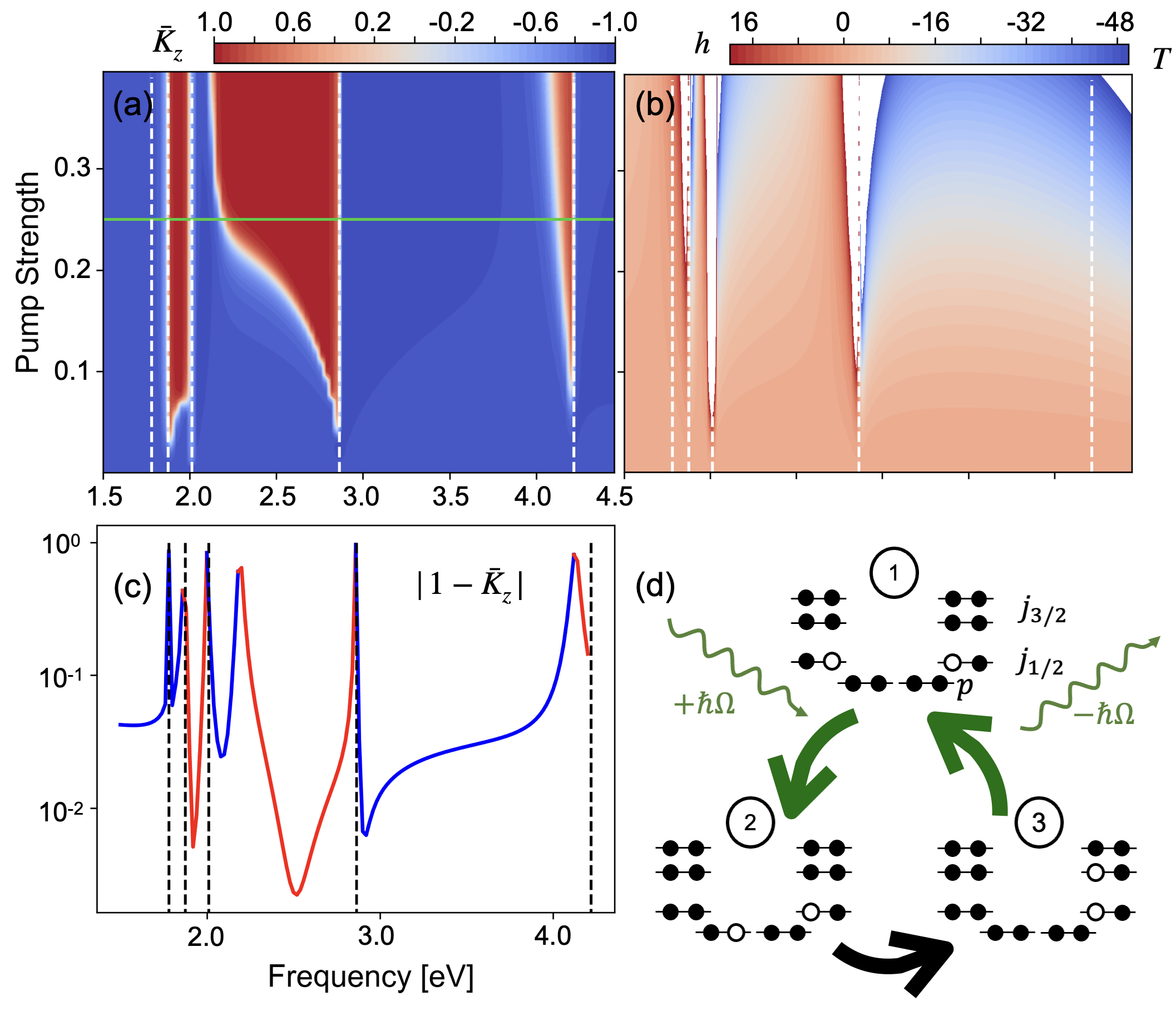}}
    \caption{
    Variation of Kitaev interaction (a) and effective magnetic field (shown in Tesla) (b) with pump parameters. Kitaev exchange $\bar{K}$ is normalized so that the magnetic coefficients output by the calculation of the model in eq. (3) are set to an energy scale of one, with $(K^2 +J^2 + \Gamma^2 + {\Gamma'}^2 + h_{ME}^2)^{1/2} \equiv 1$. Dashed white lines mark intra-$t_{2g}$ resonances. 
    (c) shows variation of $\bar{K}$ with frequency for a fixed pump frequency (green line). In a region of drive frequencies (around $\Omega = 2.5$) $\bar{K}$ approaches unity, corresponding to the Kitaev point. (d) Leading third-order time reversal symmetry breaking process responsible for the inverse Faraday effect. The chirality of the pump leads to inequivalent interference between clockwise and counter-clockwise virtual hopping of holes along TM-ligand-TM loops.} \label{fig:FloquetParameters}
\end{figure}
Here, indices $i$ and $i'$ sum over TM and ligand sites respectively, and $U, U', J_H$ denote intra-orbital, inter-orbital Coulomb interactions, and Hund's coupling for orbitals $\alpha, \beta \in \{ d_{yz}, d_{xz}, d_{xy} \}$. Furthermore, $\Delta$ parameterizes the ligand CT energy with ligand fermionic operators $\OPc{p}{i'\sigma}$, and $\lambda$ denotes the atomic SOC. 

Direct and ligand-mediated hopping of holes between neighboring Ru sites is constrained by the rotational symmetry of the octahedral plane \cite{PhysRevB.93.214431}\footnote{For simplicity, we impose $C_3$ symmetry on the hopping matrix elements, however a generalization to account for distortions of the $C2/m$ space group is straightforward.}. Here, minimal coupling between light and electrons entails a Peierls substitution $\CD{i} \C{j} \rightarrow e^{i \frac{e}{\hbar} \mathbf{r}_{ij} \cdot \mathbf{A}(t)} \CD{i} \C{j}$, where $\mathbf{r}_{ij}$ are bond vectors projected onto the octahedral plane and $\mathbf{A}(t) = A [ \sin(\Omega t), \cos(\Omega t) ]$ is the vector potential for a circularly-polarized laser with frequency $\Omega$. We henceforth denote the field strength via the dimensionless parameter $\bar{A} \equiv a_0 e \mathcal{E} / (\hbar \Omega)$, with $a_0$ and $\mathcal{E}$ the TM-TM distance and peak electric field, respectively \footnote{E.g. a dimensionless field strength of $\bar{A} = 0.25$ at a frequency $\hbar \Omega = 2.5$ eV corresponds to an electric field amplitude of $\mathcal{E} \approx 1.8 \times 10^7$ V $\cdot $ cm$^{-1}$}. Importantly, the different hopping matrix elements acquire geometrically dependent phase factors
\begin{align}
    &\hat{H}'_{ij}(t) = \sum_{\sigma} \left\{ \OPc{\mathbf{d}}{i\sigma} \hspace{-0.1cm} \cdot e^{i\mathbf{r}_{ij} \cdot \mathbf{A}(t)} \cdot \hspace{-0.1cm}  \left[\begin{array}{ccc} t_1 & t_2 & t_4 \\
    t_2 & t_1 & t_4 \\
    t_4 & t_4 & t_3 \end{array}\right] \hspace{-0.1cm} \cdot \OP{\mathbf{d}}{j\sigma} \right. \notag\\
    &+ t_{pd} \left( e^{i\mathbf{r}_{i'i} \cdot \mathbf{A}(t)}\OPc{p}{i'\sigma} \C{id_{yz}\sigma} + e^{i\mathbf{r}_{ji'} \cdot \mathbf{A}(t)}\CD{jd_{xz}\sigma} \OP{p}{i'\sigma} \right. \notag \\ &\left. \left.+ e^{i\mathbf{r}_{j'i} \cdot \mathbf{A}(t)}\OPc{p}{j'\sigma} \C{id_{yz}\sigma} + e^{i\mathbf{r}_{jj'} \cdot \mathbf{A}(t)}\CD{jd_{xz}\sigma} \OP{p}{j'\sigma} \right) + h.c. \right\},
\end{align}
with $\OPc{\mathbf{d}}{i\sigma} = \left[ \CD{id_{xz}\sigma}~ \CD{id_{yz}\sigma}~ \CD{id_{xy}\sigma} \right]$, and $i', j'$ the two ligands at the corners of edge-sharing octahedra that encase TM neighbors $i, j$ along a Z bond.

For sufficiently broad pump pulses, the Hamiltonian becomes approximately symmetric under discrete time translations $t \to t + 2\pi/\Omega$ near the pulse maximum, with $T=2\pi/\Omega$. In this case, Floquet theory permits a representation of the period-averaged transient dynamics in terms of eigenstates of an effective time-independent Hamiltonian and associated eigenenergies, which are defined modulo $\Omega$ due to lack of energy conservation. An equivalent frequency domain representation of Floquet theory seeks eigenstates of a Hamiltonian $\mathcal{H} = \sum_{mm'} (\hat{H}_{m-m'} + m\Omega \delta_{mm'}) \left|m\middle>\middle<m'\right|$ defined in a product space of the original Hilbert space and the space of periodic functions, with $\hat{H}_m = (1/T) \int_0^T dt ~e^{im\Omega t} \hat{H}(t)$ a Fourier expansion of the Hamiltonian $\hat{H}_0 + \sum_{\left<ij\right>} \hat{H}'_{ij}(t)$ \cite{supp}. Physically, this Hamiltonian describes the ``dressing'' of the original Hamiltonian with absorbed/emitted photons with frequency $\Omega$, in the semi-classical limit of a strong coherent field.

\begin{figure*}[h!t]
    \conditionalGraphics{\includegraphics[scale = 0.5,trim=0.5cm 0cm 0cm 0cm]{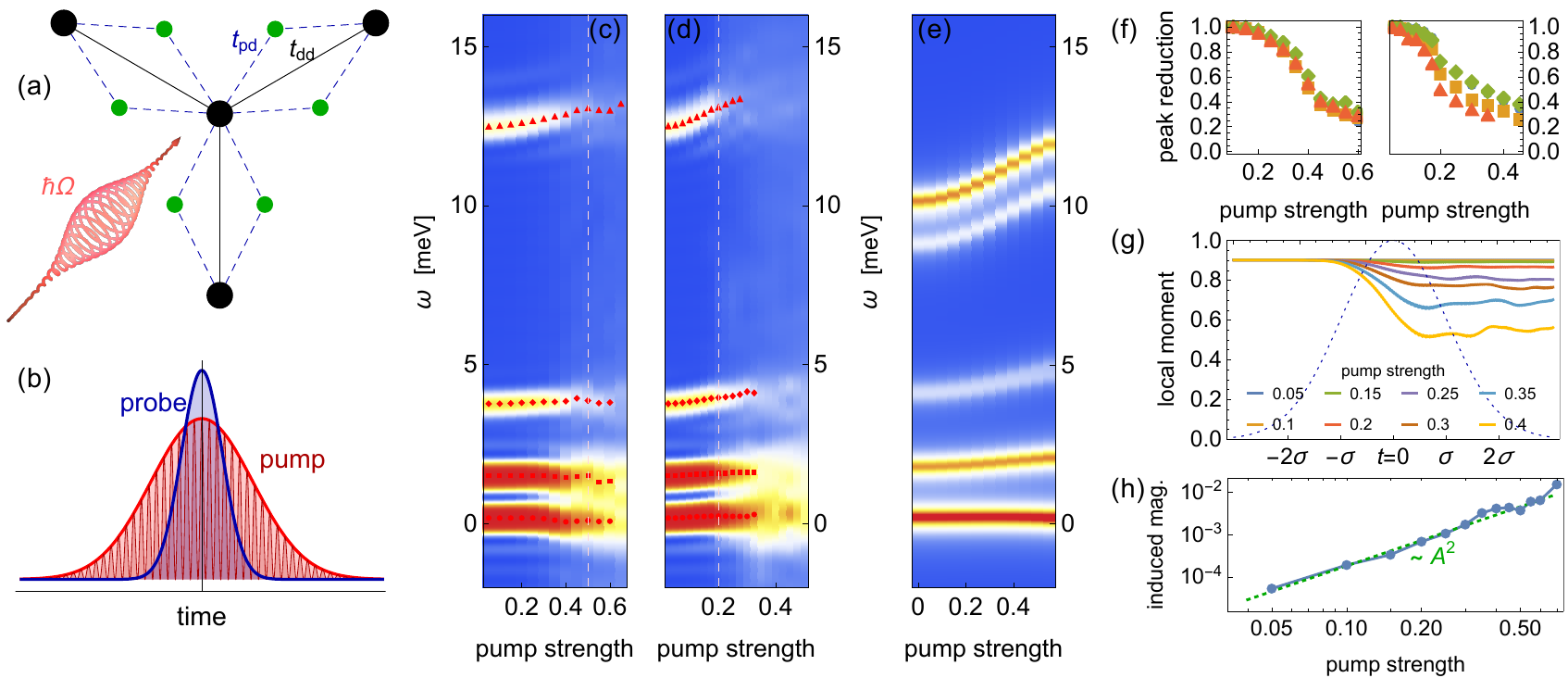}}
    \caption{
    (a) A ten-atom (18-orbital) TM-ligand cluster, consisting of a TM ion with three nearest neighbors with X, Y, Z bonds, is irradiated (b) with a broad circularly-polarized pump pulse, with the magnetic response obtained from a probe pulse centered at the pump maximum. (c), (d) Pump-probe $\mathbf{q}= 0$ dynamical spin structure factor as a function of peak pump strength $A$ for $\Omega = 0.8\textrm{eV}, 1.2\textrm{eV}$, respectively, evaluated for a 10-atom (18-orbital) cluster [inset] and probed at the peak of the Gaussian pulse. Weak fields reveal coherent modifications of the magnon spectrum, whereas the magnetic sector melts beyond a critical pump strength (dashed lines). (c) shows analogous steady-state predictions from the photo-modulated spin model, with deviations in equilibrium from the electronic model attributable to absence of charge fluctuations. (e), (f) Pump-induced suppression of the four depicted magnetic peaks [$\Omega = 0.8\textrm{eV}, 1.2\textrm{eV}$], corresponding to peak markers $\circ,\square,\diamond,\vartriangle$ in (a), (b). Melting of magnetic excitations for stronger fields coincides with time-dependent suppression of the local moment (g), which saturates for weak fields while approaching its infinite-temperature expectation value for strong pumping. (g) Transiently-induced magnetization oscillations $\langle\hat{S}^z(t)\rangle$ serve as a direct probe of the inverse Faraday effect.}
    \label{fig:dynamics}
\end{figure*}

Central to this proposal, photon dressing of electronic states entails transient modifications of the magnetic interactions which discern between ligand-mediated and direct exchange processes, as well as the generation of an effective magnetic field (an inverse Faraday effect). Both can be readily derived via a strong-coupling expansion which simultaneously eliminates charge excitations and photon absorption/emission events \cite{bukov15b,mentink15,claassen17}, and remains valid on prethermal time scales as long as the local $J=1/2$ moment persists. As ligand-mediated ($t_{pd}$) processes are essential to this description, the expansion must at minimum proceed to fourth order in $t_{pd}$. Circular polarization preserves rotations in the plane of polarization and so symmetry dictates that the resulting photo-modulated exchange interactions remain described via modified parameters $J$, $K$, $\Gamma$, $\Gamma'$, which crucially all become functions of the pump parameters $\bar{A}$, $\Omega$ \cite{supp}. Conversely, time-reversal-symmetry is broken by the field; combined with broken spin rotation symmetry due to strong SOC, this additionally necessitates the emergence of an effective photo-induced magnetic field $h_{\textrm{ME}}$. This inverse Faraday effect in a Mott insulator results purely from coupling the oscillating \textit{electric field} component of the pulse to electronic motion. Approximate $C_3$ rotation symmetry dictates that $h_{\textrm{ME}}$ must point in the [111] direction, precisely the direction required to open a gap for the non-Abelian phase of the Kitaev model. We arrive at the following prethermal magnetic Hamiltonian where all coefficients are understood to be functions of $\bar{A}, \Omega$
\begin{align}
    \hat{H} = \hspace{-0.35cm}\sum_{\langle ij \rangle _{\gamma (\alpha \beta)}} \hspace{-0.32cm}&  \mathbf{S}_i \hspace{-0.07cm} \cdot \hspace{-0.07cm} \begin{pmatrix}
    J & \Gamma & \Gamma' \\
    \Gamma & J & \Gamma' \\
    \Gamma' & \Gamma' & J + K_{\gamma}
    \end{pmatrix} \hspace{-0.12cm} \cdot \mathbf{S}_j + h_{\textrm{ME}} \sum_i \hat{\mathbf{n}}_{111}\hspace{-0.07cm} \cdot\hspace{-0.07cm} \mathbf{S}_i
\end{align}
with $ \hat{\mathbf{S}}_i^T =  \left[\begin{array}{ccc}\hat{S}^\alpha_i & \hat{S}^\beta_i & \hat{S}^\gamma_i \end{array}\right] \hspace{-0.05cm}$ and $\hat{\mathbf{n}}_{111}$ denotes the unit vector in [111] direction. Importantly, all coefficients are understood to be functions of $\bar{A}, \Omega$, and are determined numerically using a standard fourth-order strong-coupling expansion of the electronic Floquet Hamiltonian $\mathcal{H}$ defined above [see Appendix]. \footnote{Throughout this study the values of these constants used were adapted from recent \textit{ab initio} and photoemission studies: $U = 3.0, J_H = 0.45, U' = U-2J_H, \lambda = 0.15, \Delta = 5, t_1 = 0.03, t_2 = 0.03, t_3 = -0.06, t_4 = -0.02$ and $t_{pd} = -0.9$ \cite{PhysRevB.93.155143, sinn2016}.}

Results for the transient modification of Kitaev exchange (normalized to $\sqrt{K^2 + J^2 + \Gamma^2 + (\Gamma')^2 + h_{\textrm{ME}}^2}=1$) and the effective magnetic field are shown in Fig. \ref{fig:FloquetParameters}(a,b) \cite{supp}. Notably, the effect of the pump is two-fold. First, photon-dressing of the bonds leads to a bond-length-dependent renormalization of the hopping amplitudes, thereby discriminating between TM-ligand and TM-TM hopping. Second, photon absorption lowers virtual intermediate state energies. In combination, a tailored pulse can hence target the vicinity of $t_{2g}$ excitations, thereby enhancing $K$ and significantly suppressing non-Kitaev interactions [Fig. \ref{fig:FloquetParameters}(c)], nudging the material towards the QSL.

\begin{figure*}
    \conditionalGraphics{\includegraphics[scale=0.25,trim=0cm 1.0cm 0cm 0cm,clip]{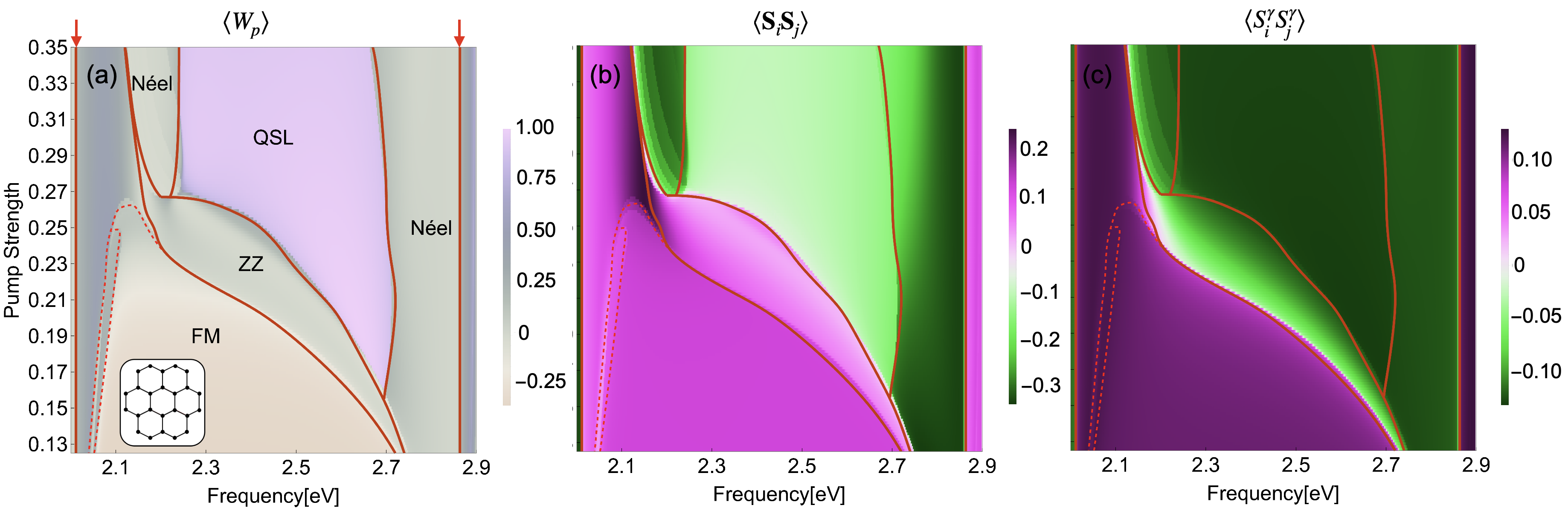}}

    \caption{Steady-state phase diagram of Floquet spin model. Color shows magnitude of $\langle W_p \rangle$ for (a), $\langle \mathbf{S}_i \mathbf{S}_j \rangle$ for (b), and $\langle S_i^{\gamma} S_j^{\gamma} \rangle$, where $\gamma$ is the Kitaev direction, for (c) that led to phase identification. In the QSL phase $\langle W_p \rangle \approx 1, \langle \mathbf{S}_i\cdot \mathbf{S}_j\rangle \approx \pm 0.131$. Red arrows denote resonances. Red lines marking phase boundaries are a guide the eye. Dashed red line marks a possible phase boundary between the ferromagnetic phase and a phase that we could not conclusively identify. Inset shows the exact diagonalization cluster used.} \label{fig:phasediagram}
\end{figure*}

\mysection{Inverse Faraday effect} The microscopic origin of the photo-induced magnetic field is readily identified as ligand-mediated \textit{third-order} virtual tunnelling processes of holes around triangles spanned by TM-ligand-TM bonds, in the presence of strong spin-orbit coupling. In equilibrium or without SOC, such processes are required to vanish in the absence of external magnetic fields, reflecting perfect destructive interference between clockwise and counter-clockwise hopping pathways. This situation changes dramatically in the driven system. Here, a hole can virtually absorb and emit a photon upon hopping along the non-collinear $d-d$ and $p-d$ bonds of a triangle [Fig. \ref{fig:FloquetParameters}(d)], thereby picking up an overall phase to avert cancellation with the directionally-reversed ``partner'' process. This phase in turn depends on the ligand bond angle.


As the contribution to individual TM sites must follow from a sum of processes along X, Y, Z bonds, approximate $C_3$ rotation symmetry entails that the total induced magnetic field points in the [111] direction. Its quantitative dependence on pump strength and frequency is depicted in Fig. \ref{fig:FloquetParameters}(b), including up to fourth-order hopping processes. While the magnetic field experiences a series of sign changes that coincide with intermediate-state resonances, it is instructive to approximate the photo-induced magnetic field for low-frequency pumping near the first resonance
\begin{align}
    h_{\textrm{ME}}(A,\Omega) \approx \frac{A^2 ~\sin(\phi) ~t_{pd}^2 ~(t_1 - t_3) }{36 r_{pd} (U - 3J_H - \frac{3}{2}\lambda - \Omega)}  \left( \frac{1}{\Delta} - \frac{1}{\Delta + \Omega} \right)
\end{align}
Here, $\phi$ denotes the bond angle between TM-TM and TM-ligand bonds projected into the plane of polarization, and $r_{pd}$ is the relative TM-ligand bond length compared to the direct TM-TM distance.

\mysection{Spectroscopic Signatures} 

Armed with predictions for light-induced magnetic interactions, we now turn to consequences for the transient magnetic state. There are two prescient questions: (1) Can the coherent modification of the local exchange interactions predicted by the Floquet picture be observed in time-resolved pump probe experiments, where energy absorption is an obstacle? (2) Does the the prethermal steady state phase diagram includes a non-equilibrium Kitaev QSL phase that can be reached with optical means?  

To address the former, we employ large-scale numerical simulations of an electronic 18-orbital ``Ru\textsubscript{4}Cl\textsubscript{6}'' cluster [Fig. \ref{fig:dynamics} (a)] to study a non-equilibrium generalization of the dynamical $q=0$ spin structure factor \cite{wang17}
\begin{align}
    S(\omega,t_p) = \int dt dt' e^{i\omega(t-t')} f_{t_p}(t) f_{t_p}(t') \left< \hat{S}^z(t) \hat{S}^z(t') \right>  \label{eq:dynamicalSpinResponse}
\end{align}
where $\hat{S}^z = (1/2)\sum_{n\alpha} (\ND{n\alpha\uparrow} - \ND{n\alpha\downarrow})$, and $f_{t_p}(t) = e^{-(t-t_p)^2 / 2\sigma_p}$ are Gaussian probe shape functions with width $\sigma_p$, centered at the probe time $t_p$. The two-time spin correlation function is calculated from time evolution of the exact electronic many-body wave function \cite{supp}. This permits a direct study of the resulting photo-induced magnon spectrum for realistic pulse envelopes, to assess the competition between coherent control and heating.

Starting from the equilibrium ground state, a wide ($\sim$ 1 ps) Gaussian pulse with circular polarization irradiates the system and a second Gaussian probe pulse measures the magnetic response [Fig. \ref{fig:dynamics} (b)]. To assess the maximal change of the magnetic excitation spectrum, we probe the dynamical spin response [Eq. (\ref{eq:dynamicalSpinResponse})] at the peak of the pump pulse. Fig. \ref{fig:dynamics}(c) and (d) depict the pump strength dependence for two pump frequencies $\Omega=0.8$eV and $\Omega=1.2$eV, respectively. While energy absorption from the pump would naively be expected to supress magnetism, the observed response exhibits a \textit{coherent} hardening of higher-energy magnetic excitations at weak pump strengths. This behavior corresponds to a prethermal enhancement of  magnetic exchange interactions as predicted above. Fig. \ref{fig:dynamics}(e) shows qualitatively-analogous steady-state predictions for the Floquet spin Hamiltonian [Fig. \ref{fig:FloquetParameters}] as a function of pump strength at $\Omega=0.8$eV, with equilibrium discrepancies of excitation energies attributable to charge fluctuations that are amiss in a purely magnetic description. Beyond a critical pump strength, the magnetic peaks are rapidly suppressed due to heating [Fig. \ref{fig:dynamics}(f)]. This melting of the magnetic sector at $A \approx 0.4$ ($\approx 0.2$ for $\Omega=1.2$eV) coincides with a time-dependent suppression of magnetic moments [Fig. \ref{fig:dynamics}(g)], which approach their infinite-temperature expectation values beyond the critical pump strength. Finally, while the ground state in equilibrium has zero net magnetization, the inverse Faraday effect contribution can also be individually probed from time-dependent induced magnetization oscillations. Fig. \ref{fig:dynamics}(h) depicts the peak magnetization $\max \left|\left< S^z(t) \right>\right|$ as a function of pump strength near the pump maximum, which matches the $\sim A^2$ scaling as expected from Floquet predictions.


\mysection{Floquet phase diagram} To assess whether the predicted optical control of the magnetic interactions can nudge the system towards a QSL phase, we now turn to studying the transient phase diagram via exact diagonalization of the photo-modulated \textit{magnetic} Hamiltonian for a larger 24-site cluster. Physically, this corresponds to the limit in which the equilibrium state can transforms close to adiabatically to the corresponding prethermal Floquet state for slow switch-on of the drive, effectively identifying phase boundaries that remain inaccessible in equilibrium but can be reached by driving, and effectively delineates reachable out-of-equilibrium phase transitions for appropriately tailored pulses. The steady-state phase diagram is depicted in Fig. \ref{fig:phasediagram} as a function of pump strength and frequency, with the particular region of pump parameters shown corresponding to a region where the Kitaev interaction is predicted to be sufficiently dominant as to stabilize a QSL [Fig. \ref{fig:FloquetParameters}(c)]. Phase boundaries were determined from singular features in $-\frac{\partial^2 E}{\partial \omega^2}$, and phases were identified from the value of the plaquette flux operator $W_p = 2^6 S_i^x S_j^y S_k^z S_l^x S_m^y S_n^z$ ($\pm 1$ for the QSL) [\ref{fig:phasediagram}(a)] and spin correlation functions [\ref{fig:phasediagram}(b)]. 
Notably, the pump parameters provide handles to traversing the $JK\Gamma \Gamma'$ phase diagram, suggesting that the equilibrium zigzag (ZZ) phase of RuCl\textsubscript{3} could be pushed into a number of proximal phases.

There are two striking features of the identified spin liquid phase. First, although RuCl\textsubscript{3} is proximate to a ferromagnetic-$K$ QSL, we find here that the optimal choice of pump parameters leaves the sign of $K$ transiently inverted. Second, in contrast to the equilibrium phase diagram \cite{chaloupka12,rau14,winterNComm2017,supp}, the identified QSL occupies a robust and wide range of the parameter space. This can be attributed precisely to the dynamical inverse Faraday effect that gaps out the QSL, consistent with recent theoretical studies on QSL stability in a magnetic field \cite{Gordon2019} \footnote{Furthermore, a small but nonzero $\Gamma'$ is also purported to stabilize the QSL phase \cite{PhysRevB.99.224409}.}. Finally, we remark on the dashed red lines shown in [\ref{fig:phasediagram}]. In the region where $2 < A < 2.1$, the spin exchange coefficents derived from the Floquet calculation take on values which place the system in a phase that is possibly ferromagnetic, incommensurate spiral or zigzag. However, there is not widespread agreement on what the resulting phase is and we were unable to readily identify it \cite{rau14, winterNComm2017}.

\mysection{Discussion}
We have shown that optical irradiation of $\alpha$-RuCl\textsubscript{3} can grant a two-fold handle, to transiently alter competing magnetic interactions and induce an effective magnetic field, thereby coherently manipulating the magnetic state. Predicted spectroscopic signatures of coherent control presented above are readily accessible in time-resolved experimental probes of the magnetic excitation spectrum.

A lingering question regards possible time resolved signatures of spin liquid behavior out of equilibrium. Here, the central conclusion drawn from the steady-state phase diagram established in this work is that the equilibrium zigzag state can be ``nudged'' towards the QSL regime using an appropriately tailored pulse. Therefore, an intriguing consequence is the possibility for signatures of phase boundaries in the spectroscopic response. For instance, we speculate that Kerr and Faraday rotation, while generally yielding a response upon circularly-polarized irradiation, may be effective at detecting changes corresponding phase boundaries between photo-induced phases (e.g. AFM to QSL). The pump strengths presented here should be accessible experimentally \cite{kawakami}. We also note that understanding the pulse shape dependence of such prethermal non-equilibrium phase transitions remains an interesting open problem \cite{kennes18}.

Furthermore, while energy absorption from the pump melts magnetic order at higher pump strengths as shown above, a general omission in Floquet engineering studies of correlated materials is the lack of energy dissipation to the lattice or other degrees of freedom. Here, an important open question concerns the role of energy absorption, dissipation and decoherence, in stabilizing or destabilizing transiently-engineered non-equilibrium states.

Finally, the electronic model for RuCl\textsubscript{3} remains under considerable debate, with a minimal description utilized in this work neglecting the $e_g$ orbitals of the TM as well as longer-ranged exchange processes beyond nearest neighbors. The former contribute to magnetic exchange \cite{chaloupka12}; however \textit{ab initio} studies of RuCl\textsubscript{3} that neglect $t_{2g} - e_g$ mixing have successfully captured magnetic properties in equilibrium \cite{PhysRevB.93.214431, PhysRevB.93.155143};  inclusion of $e_g$ orbitals is hence not expected to significantly alter our results as long as direct resonances are avoided. For the latter, several studies remarked on the importance of third-neighbor interactions in stabilizing the equilibrium zigzag phase \cite{PhysRevB.93.214431,winterNComm2017,sears20}. However, it is unclear whether the $\Gamma'$ interaction or the $J_3$ interaction is more important in this regard \cite{sears20}. Their relative importance away from equilibrium hence remains an open question for future work. 

Importantly, the predicted mechanism readily generalizes to other spin orbit coupled magnets, including the honeycomb Kitaev materials such as Na\textsubscript{2}IrO\textsubscript{3} and $\alpha$-Li\textsubscript{2}IrO\textsubscript{3}. Here, while details of the relative competition of magnetic parameters can differ dramatically, a central thread is the possibility to leverage the added materials complexity of ligand-mediated interactions to provide a knob to both affect the interplay of anisotropic exchange and induced magnetic fields in a controlled manner, to steer the magnetic state. We therefore expect that our analysis can be readily applied to a variety of frustrated magnetic systems.

\acknowledgements{We would like to thank Liang Wu and James McIver for helpful comments and discussions. We are grateful for hospitality and support from the Center for Computational Quantum Physics at the Flatiron Institute, a division of the Simons Foundation, where part of this work was performed.}

\newpage
\appendix
\section{Models and Methods}
\subsection{Floquet Theory}
In this section we will provide a very brief overview of Floquet theory. Similar to Bloch's theorem which provides an ansatz for the wave function in the presence of discrete spatial translational symmetry, Floquet theory provides an ansatz for the wave function when there is discrete time translational symmetry \cite{oka18}.
\begin{align}
    \hat{H}(t) = \hat{H}(t+T)
\end{align}
Consider then the unitary time evolution operator $U(t+T,t)$ which evolves a state from $t$ to $t+T$. The eigenstates of this operator are given by
\begin{align}
    \hat{U}(t+T,t)\ket{\phi_n} = \lambda_n \ket{\phi_n}
\end{align}
Because the time evolution operator is a unitary operator, the eigenvalues must obey the relation
\begin{align}
    \lambda_n \lambda_n^* = 1
\end{align}
This is generally satisfied by
\begin{align}
    \lambda_n = e^{-i\epsilon_n T}
\end{align}
where $\epsilon_n$ is referred to as the Floquet quasi-energy and is uniquely defined on the interval $(\pi/T, \pi/T]$. As the time evolution operator solves the Schrodinger equation and commutes with the Hamiltonian, the eigenstates of $U$ are also solutions to the time-dependent Schrodinger equation. Using the definition of $U(t+T,t)$ we have
\begin{align}
    \hat{U}(t+T)\ket{\phi_n(t)} = \ket{\phi_n(t+T)} = e^{-i\epsilon_nT}\ket{\phi_n(t)}
\end{align}
From this, we see that the states are periodic up to the factor involving the quasi-energy. This allows us to express the state as a Fourier series giving us the general Floquet ansatz
\begin{align}
    \ket{\phi_n} = e^{-i\epsilon_nT}\sum_m \ket{m}e^{im \frac{2\pi}{T} t}
\end{align}
Similarly, we can express the periodic Hamiltonian also in Fourier form. This is important to removing the time degree of freedom and replacing it with a frequency degree of freedom in the Schrodinger equation.

\subsection{Floquet Kanamori Model}

The Peierl's substitution is given by
\begin{align}
    c_i^{\dagger}c_j \rightarrow e^{i\mathbf{r}_{ij} \cdot \mathbf{A}(t)}c_i^{\dagger}c_j.
\end{align}
The orientation of the field we study is given by
\begin{align}
    \mathbf{A}(t) = A(t)\begin{pmatrix} \cos(\omega t) & \sin(\omega t) \end{pmatrix}^T.
\end{align}
At steady state, $A(t)$ is a constant. In the presence of the photon field, the multi-orbital Hubbard model is given by
\begin{align}
    \hat{H} = \hat{H}_0 + \hat{H}'_{ij}(t).
\end{align}
The Kanamori Hamiltonian is unchanged by the drive and is given in (2) in the main text but the hopping Hamiltonian now contains the aformentioned Peierl's substitutions:  
\begin{align}
    &\hat{H}'_{ij}(t) = \sum_{\sigma} \left\{ \left[ \CD{id_{xz}\sigma}~ \CD{id_{yz}\sigma}~ \CD{id_{xy}\sigma} \right] \right. \hspace{-0.1cm} \notag\\
    &\cdot e^{i\mathbf{r}_{ij} \cdot \mathbf{A}(t)} \cdot \hspace{-0.1cm}  \left[\begin{array}{ccc} t_1 & t_2 & t_4 \\
    t_2 & t_1 & t_4 \\
    t_4 & t_4 & t_3 \end{array}\right] \hspace{-0.1cm} \cdot \hspace{-0.1cm}  \left[\begin{array}{c} \C{jd_{xz}\sigma} \\ \C{jd_{yz}\sigma} \\ \C{jd_{xy}\sigma} \end{array}\right] \notag\\
    &+ t_{pd} \left( e^{i\mathbf{r}_{i'i} \cdot \mathbf{A}(t)}\OPc{p}{i'\sigma} \C{id_{yz}\sigma} + e^{i\mathbf{r}_{ji'} \cdot \mathbf{A}(t)}\CD{jd_{xz}\sigma} \OP{p}{i'\sigma} \right. \notag \\ &\left. \left.+ e^{i\mathbf{r}_{j'i} \cdot \mathbf{A}(t)}\OPc{p}{j'\sigma} \C{id_{yz}\sigma} + e^{i\mathbf{r}_{jj'} \cdot \mathbf{A}(t)}\CD{jd_{xz}\sigma} \OP{p}{j'\sigma} \right) + h.c. \right\}
\end{align}
If we plug in the Floquet ansatz into the Schrodinger equation, we get
\begin{align}
    \sum_m (\epsilon_n - m\omega) e^{im\omega t} \ket{m} &= \left[ \hat{H}'_{ij}(t) + \hat{H}_K \right]\sum_{m'} e^{im'\omega t} \ket{m'}
\end{align}
$\hat{H}'_{ij}$ contains the periodic driving term and so can be expanded as a Fourier series. The time degree of freedom is eliminated in favor of the pump frequency by expressing $\hat{H}'_{ij}$ as
\begin{align}
    \hat{H}_{m} = \frac{2\pi}{\Omega}\int_0^{2\pi/\Omega} dt \quad e^{-im\Omega t}\hat{H}'_{ij}
\end{align}
From here, it is straightforward to recover the multi-orbital Floquet-Hubbard model. This Hamiltonian should produce the spectrum of Floquet quasi-energy spectrum.
\begin{align}
    &\hat{H}'_{FH} = \sum_{\sigma} \left\{ \left[ \CD{id_{xz}\sigma}~ \CD{id_{yz}\sigma}~ \CD{id_{xy}\sigma} \right] \right. \hspace{-0.1cm} \cdot \hspace{-0.1cm} \notag\\
    &\mathcal{J}_{m-m'}(\bar{A}(a))e^{i(m-m')arg\mathbf{r}_{ij}} \cdot \hspace{-0.1cm}  \left[\begin{array}{ccc} t_1 & t_2 & t_4 \\
    t_2 & t_1 & t_4 \\
    t_4 & t_4 & t_3 \end{array}\right] \hspace{-0.1cm} \cdot \hspace{-0.1cm}  \left[\begin{array}{c} \C{jd_{xz}\sigma} \\ \C{jd_{yz}\sigma} \\ \C{jd_{xy}\sigma} \end{array}\right] \notag\\
    &+ t_{pd} \mathcal{J}_{m-m'} (\bar{A}(a)) \left( e^{i(m-m')arg\mathbf{r}_{i'i}}\OPc{p}{i'\sigma} \C{id_{yz}\sigma} \right. \notag \\ &+ e^{i(m-m')arg\mathbf{r}_{ji'}}\CD{jd_{xz}\sigma} \OP{p}{i'\sigma}+ e^{i(m-m')arg\mathbf{r}_{j'i}}\OPc{p}{j'\sigma} \C{id_{yz}\sigma}  \notag \\ &\left. \left.+ e^{i(m-m')arg\mathbf{r}_{j'j}}\CD{jd_{xz}\sigma} \OP{p}{j'\sigma} \right) + h.c. \right\} \otimes \ket{m'}\bra{m} \notag \\
    &+ \hat{H}_0 \otimes \mathds{1} + \sum_m m\Omega \mathds{1} \otimes \ket{m}\bra{m}
\end{align}
where $a,b \in \{d_{yz}, d_{xz}, d_{xy} \}$,  $\bar{A} = A\hbar / (ea)$, describing the dimensionless field strength at the pump plateau, $A$ being the steady state value of $A(t)$ and $a$ the distance between sites (note this is different depending on whether the electron is hopping to a $p$ orbital or a transition metal). $\mathcal{J}_n (\cdot)$ is the Bessel function of the first kind and is a result of applying the Jacobi-Anger identity to the exponential containing a trigonometric term. 

We performed the complete perturbation theory of the periodically driven model numerically to determine magnetic interactions by using exact diagonalization on a two site Ru\textsubscript{2}Cl\textsubscript{2} electronic model. This allowed the extraction of the nearest neighbor magnetic exchange coefficients and their dependence upon the pump parameters.

\subsection{Spectroscopic Response \label{app:spectroscopy}}

We characterize the modification of the magnetic excitation spectrum via large-scale numerical simulations of the two-time-resolved magnetic response of a 18-orbital cluster of four transition metal sites ($t_{2g}$ manifold) and six ligand sites ($p$-orbital). The main text illustrates these results via a straightforward non-equilibrium generalization of the dynamical $q=0$ spin structure factor \cite{wang17}
\begin{align}
    S(\omega,t_p) = \int dt dt' e^{i\omega(t-t')} f_{t_p}(t) f_{t_p}(t') \left< \hat{S}^z(t) \hat{S}^z(t') \right>
\end{align}
where $\hat{S}^z = (1/2)\sum_{n\alpha} (\ND{n\alpha\uparrow} - \ND{n\alpha\downarrow})$, and $f_{t_p}(t) = e^{-(t-t_p)^2 / 2\sigma_p}$ are Gaussian probe shape functions with width $\sigma_p$, centered at the probe time $t_p$.

The spin correlation functions $\left< \hat{S}^z(t) \hat{S}^z(t') \right>$ are evaluated on a two-time grid using using highly-parallelized time evolution of the exact many-body wave function, via a Krylov-subspace representation of the pump-dependent time evolution operator with small time steps. As the main focus lies on the magnetic response at $meV$ energy scales, sampling is performed on a $9690 \times 9690$ two-time grid with spacing $\delta t = 0.658 {\rm fs}$. This corresponds to two-time evolutions up to 96900 steps, using a time spacing $\delta t / 10$ with 10 Lanczos iterations per time step. These parameters suffice to sharply resolve magnetic features before the onset of heating, and reproduce the equilibrium magnetic response (calculated using a Lanczos representation of the resolvent) in the absence of the pump.

\subsection{Steady-State Phase Diagram}
To characterize the phase behavior of the non-equilibrium system, we performed exact diagonalization calculations on a 24-site cluster that has commonly been used previously to study the honeycomb magnetic model, as it contains the full point-group symmetry of the honeycomb lattice and in addition, it contains the high symmetry points of the Brillouin zone. Cartoon schematics of this cluster as well as its supported phases are shown in appendix D. We use periodic boundary conditions with lattice vectors defined as $\mathbf{R}_1 = a(5\sqrt{3}/2, 9/2)$ and $\mathbf{R}_2 = a(5\sqrt{3}/2,0)$. The lowest lying eigenvalues and eigenvectors were determined using the Lanczos method.

Magnetic phases were identified using the calculated observables shown in figure 3 of the main text, boundaries drawn according to singularities in $\partial ^2 E$ \cite{PhysRevB.95.024426}. In certain regions in the pump parameter phase space, there exists changes in the observables with a boundary marked by a singularity in $\partial ^2 E$, but the phases calculated across both sides of this boundary were the same. These were due to finite size effects of the small cluster. In figure 3, only the boundaries marking distinctive phases are shown. As a check, we reproduced the full phase diagram of the $JK\Gamma$ model from \cite{PhysRevB.95.024426, rau14, winterNComm2017} in Fig. \ref{fig:phasediagramJKG}.

\section{Behavior of Other Photomodified Magnetic Interactions}
The main text showed the photomodified normalized Kitaev interaction and unnormalized effective magnetic field in units of Tesla. In normalized dimensionless units, $K \approx 1$ automatically means $J, \Gamma, \Gamma' \approx 0$. Here we show the other coefficients. In addition, we show the value of the normalization factor which is effectively the energy scale. Although the normalized model may predict phases very close to resonant peaks, if the energy scale in that regime is too large one would expect excessive heating. This is why in the main text, we explore the phase diagram far off-resonant but where there is still a large change in the exchange coefficients. 
\begin{figure*}[h]
\conditionalGraphics{\includegraphics[scale=0.35]{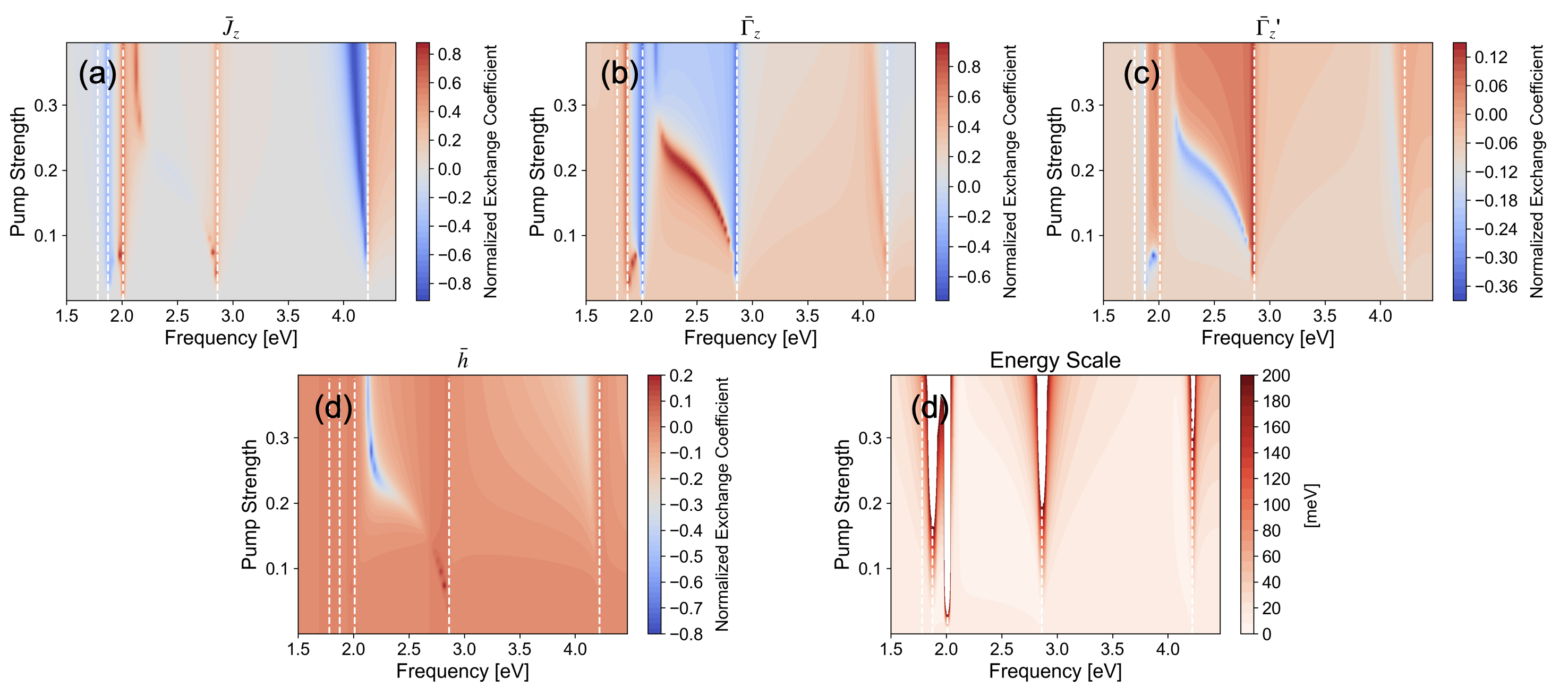}}
\caption{\captiontitle{Pump parameter dependence of $\bar{J},\bar{\Gamma}, \bar{\Gamma'}, \bar{h}$ and the normalization factor.} (a)-(d) show the dependence of isotropic spin exchange $J$, off-diagonal exchanges $\Gamma$, $\Gamma'$ and the photo-induced magnetic field, as a function of the pump parameters, and normalized to unity $\sqrt{K^2 + J^2 + \Gamma^2 + \Gamma'{}^2 + h^2 = 1}$. (e) depicts the normalization factor that sets the overall energy scale.}
\end{figure*}

\section{Floquet Engineered Magnetic Model Behavior in Other Frequency Regimes}
\begin{figure*}[h]
\conditionalGraphics{\includegraphics[scale=0.35]{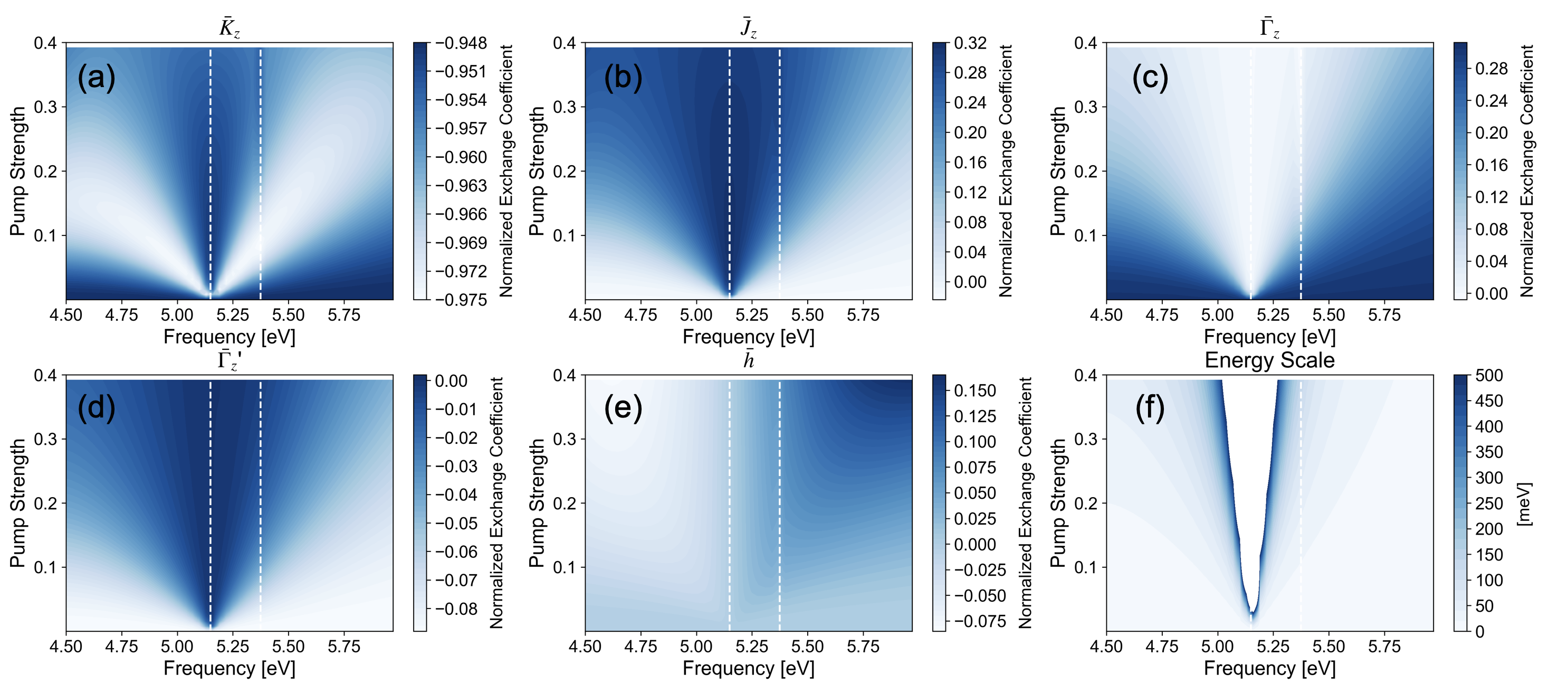}}
\caption{\captiontitle{Pump parameter dependence of Spin-Exchange Coefficients at Higher Frequency} } \label{fig:magneticParamsHighFreq}
\end{figure*}

Here we show the behavior of the photo-modified magnetic model in other frequency regimes. Fig. \ref{fig:magneticParamsHighFreq} shows the behavior close to the ligand single-occupancy regime. Although there is a quite a bit of interesting behavior in this regime, the energy scale here diverges quite rapidly. This is shown in Fig. \ref{fig:magneticParamsHighFreq}(f). We restricted the color scale to show contrast up to 500 meV but the scale exceeds this value at some distance away from the resonant frequencies. 

\begin{figure*}[h]
\conditionalGraphics{\includegraphics[scale=0.35]{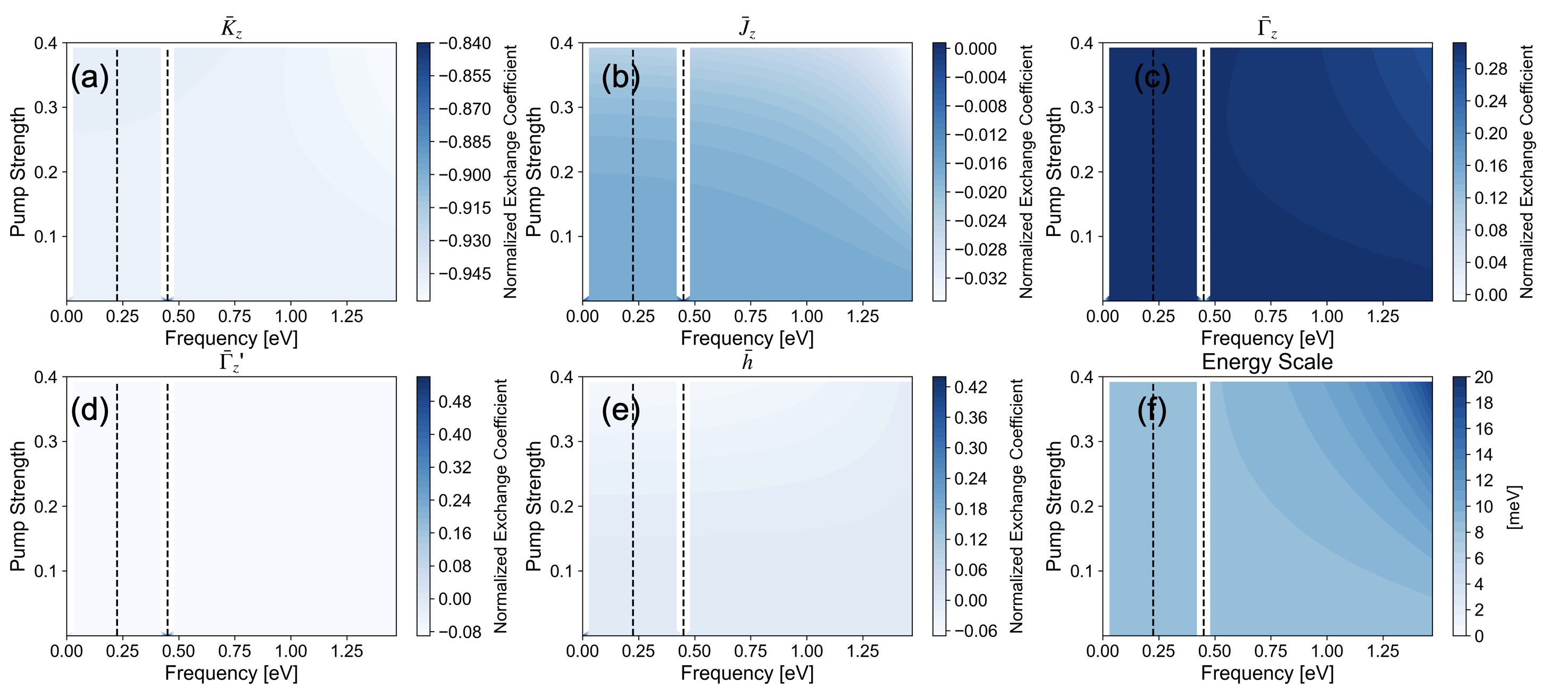}}
\caption{\captiontitle{Pump parameter dependence of Spin-Exchange Coefficients at Lower Frequency}} \label{fig:magneticParamsLowFreq}
\end{figure*}

We also show the behavior in the lower frequency regime [Fig. \ref{fig:magneticParamsLowFreq}]. This range of frequencies contains the low energy spin-orbit coupling resonances. Here modulations of the magnetic interactions are small; the energy scale is also smaller. This regime also contains the magnetic model behavior at the frequencies at which we computed spectroscopic signatures in figure 2 in the main text. 

\section{Magnetic Model}
In the main text, we studied the $JK\Gamma \Gamma'$ model given by Eq. 3 (main text). This model arises upon treating the kinetic hopping term as a perturbation to the Kanamori Hamiltonian. The general nearest neighbor magnetic model takes the form
\begin{align}
    \mathcal{H} = \sum_{\langle ij \rangle} J_{ij} \mathbf{S}_i \cdot \mathbf{S}_j + \mathbf{D}_{ij} \mathbf{S}_i \times \mathbf{S}_j + \mathbf{S}_i \cdot \mathbf{T}_{ij} \cdot \mathbf{S}_j
\end{align}
where $J_{ij}$ is the Heisenberg exchange, $\mathbf{D}_{ij}$ describes the Dzyaloshinskii-Moriya (DM) interaction and $\mathbf{T}_{ij}$ describes the pseudo-dipolar interaction \cite{PhysRevB.93.214431}. The (DM) interaction exists only in the presence of inversion symmetry breaking and so in the models presented in this study, it was 0. Along each bond, the presence of inversion symmetry simplifies the pseudo-dipolar interaction and leads to the $JK\Gamma \Gamma'$ model with interaction matrices given by
\begin{align}
    \begin{pmatrix}
    J_n^z & \Gamma_n^z & \Gamma_n^{'z}\\
    \Gamma_n^z & J_n^z & \Gamma_n^{'z}\\
    \Gamma_n^{'z} & \Gamma_n^{'z} & J_n^z + K_n^z
    \end{pmatrix} \\
    \begin{pmatrix}
    J_n^{xy} + K_n^{xy} & \Gamma_n^{'xy} + \zeta_n & \Gamma_n^{'xy} - \zeta_n\\
    \Gamma_n^{'xy} + \zeta_n & J_n^x & \Gamma_n^x\\
    \Gamma_n^{'xy} - \zeta_n & \Gamma_n^{xy} & J_n^{xy}
    \end{pmatrix}\\
    \begin{pmatrix}
    J_n^{xy} & \Gamma_n^{'xy} + \zeta_n & \Gamma_n^{xy}\\
    \Gamma_n^{'xy} + \zeta_n & J_n^{xy} + K_n^{xy} & \Gamma_n^{'xy} - \zeta_n\\
    \Gamma_n^{'xy} & \Gamma_n^{'xy} - \zeta_n & J_n^{xy}
    \end{pmatrix}
\end{align}
We note here that the Z bond in this treatment has further symmetry than the X or Y bonds and therefore takes on a simpler form but if C3 symmetry is assumed then $J_n^z = J_n^{xy}, K_n^z = K_n^{xy}, \zeta_n = 0, \Gamma_n^z = \Gamma_n^{xy},\Gamma_n^{'z} = \Gamma_n^{'xy}$. If one further assumes octahedral symmetry (which we do not in this study), this takes $t_4 = 0$ and also results in $\Gamma' = 0$. In our calculations however, we found that the $\Gamma'$ interaction was critical to stabilizing zigzag magnetic order in RuCl\textsubscript{3} in equilibrium.

In equilibrium, the exchange coefficients calculated by fourth order perturbation theory from the multi-orbital Floquet Hubbard model (equation (1)) were given by: $J = -0.03, K = -8.21, \Gamma = 2.48, \Gamma' = -0.74$. These validity of these values were supported by a number of other studies, and namely we note the large ferromagnetic $K$ and antiferromagnetic $\Gamma$ as well as the small ferromagnetic $J$ \cite{PhysRevB.93.214431, Winter_2017, PhysRevB.93.155143}. When simulated on the 24-site cluster, these exchange coefficients indeed produced the desired equilibrium zigzag phase.

\begin{figure*}[h]
    \hspace*{\fill}%
    \subfloat[N\'eel Antiferromagnet][]{\conditionalGraphics{\includegraphics[scale=0.2]{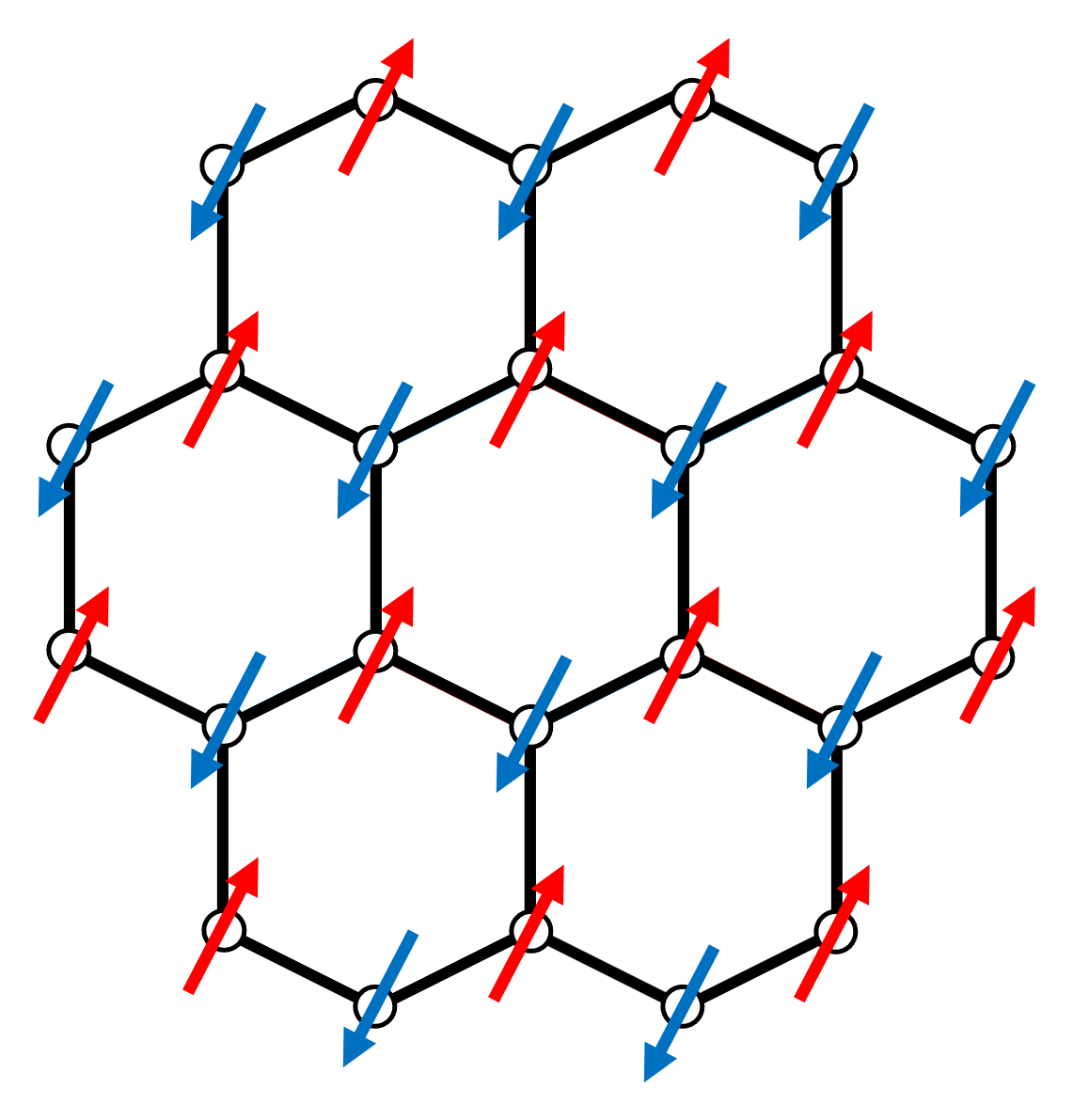}}}
    \hfill
    \subfloat[Zigzag][]{\conditionalGraphics{\includegraphics[scale=0.2]{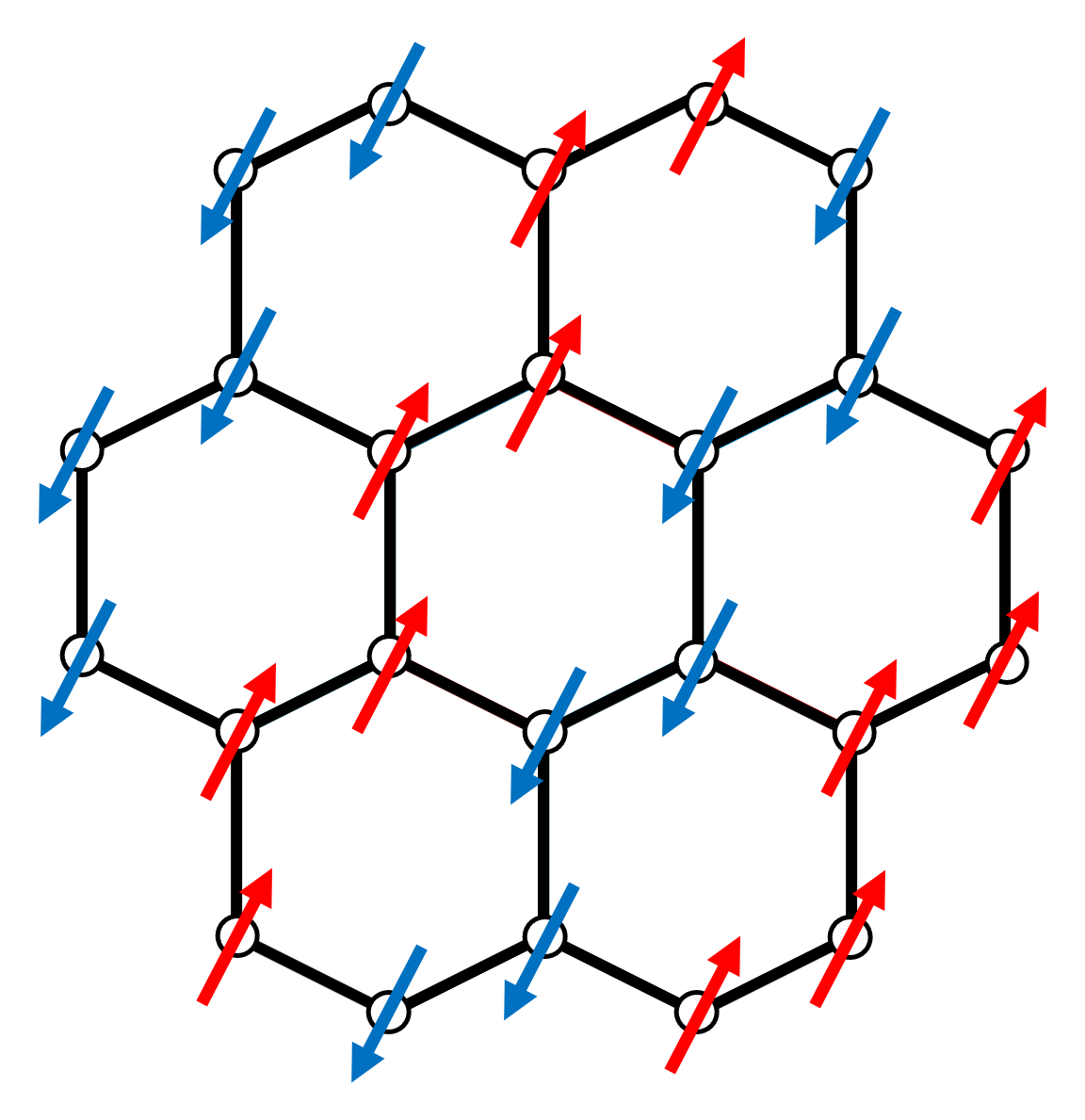}}}
    \hfill
    \subfloat[Stripy][]{\conditionalGraphics{\includegraphics[scale=0.2]{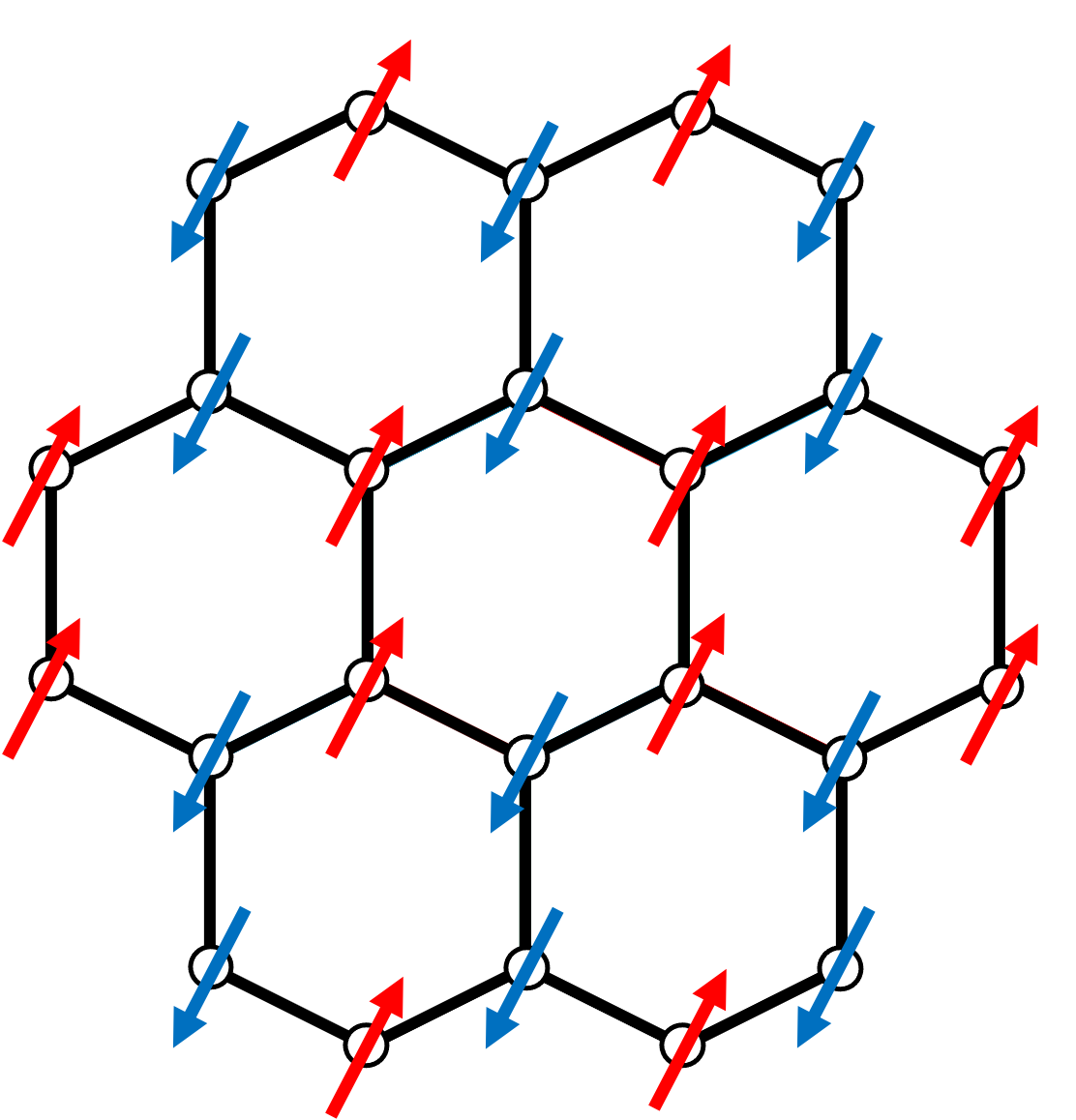}}}
    \hspace*{\fill}
    \\~\\
    \hspace*{\fill}%
    \subfloat[Ferromagnet][]{\conditionalGraphics{\includegraphics[scale=0.2]{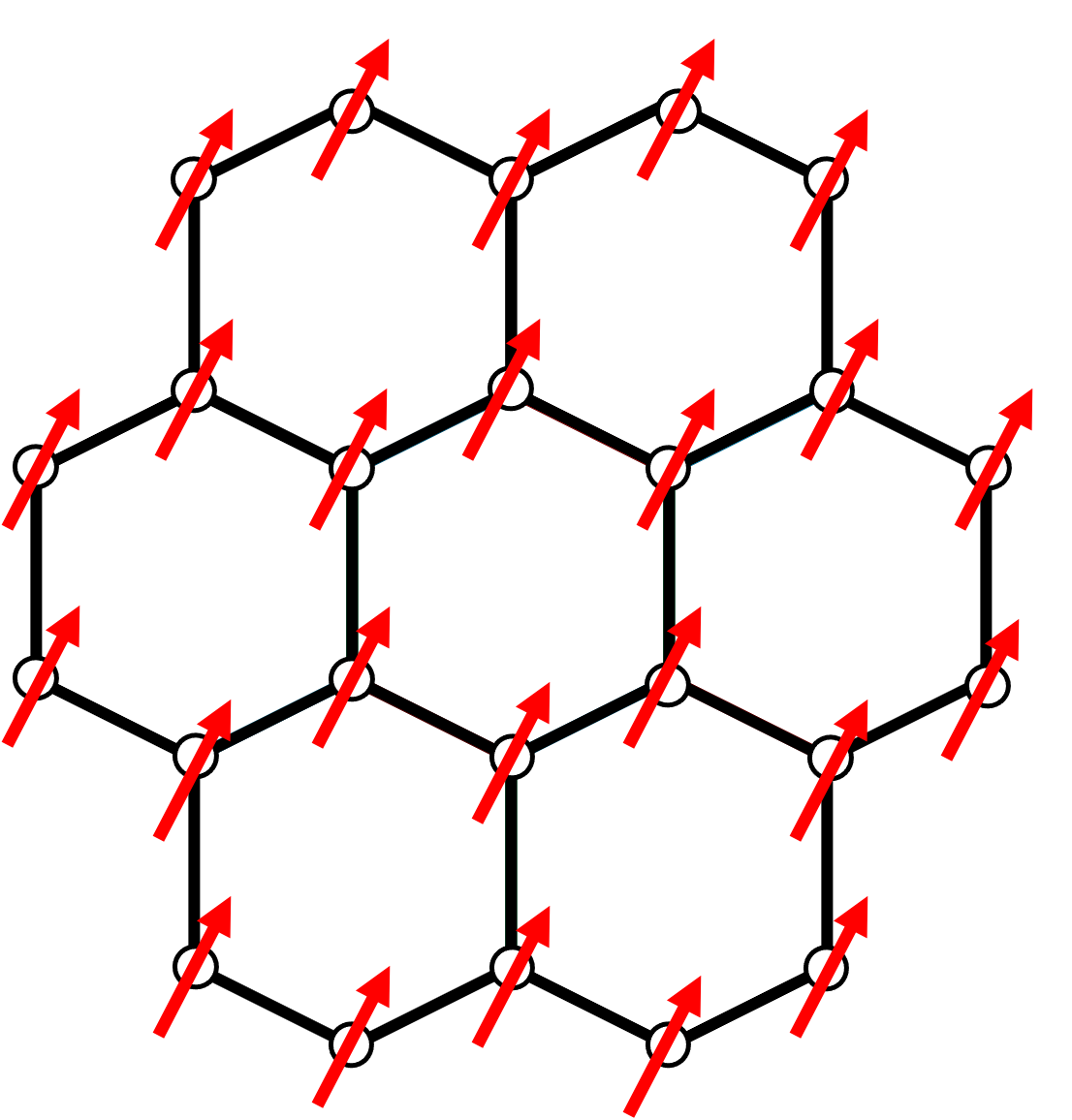}}}
    \hfill
    \subfloat[$120^{\circ}$][]{\conditionalGraphics{\includegraphics[scale=0.2]{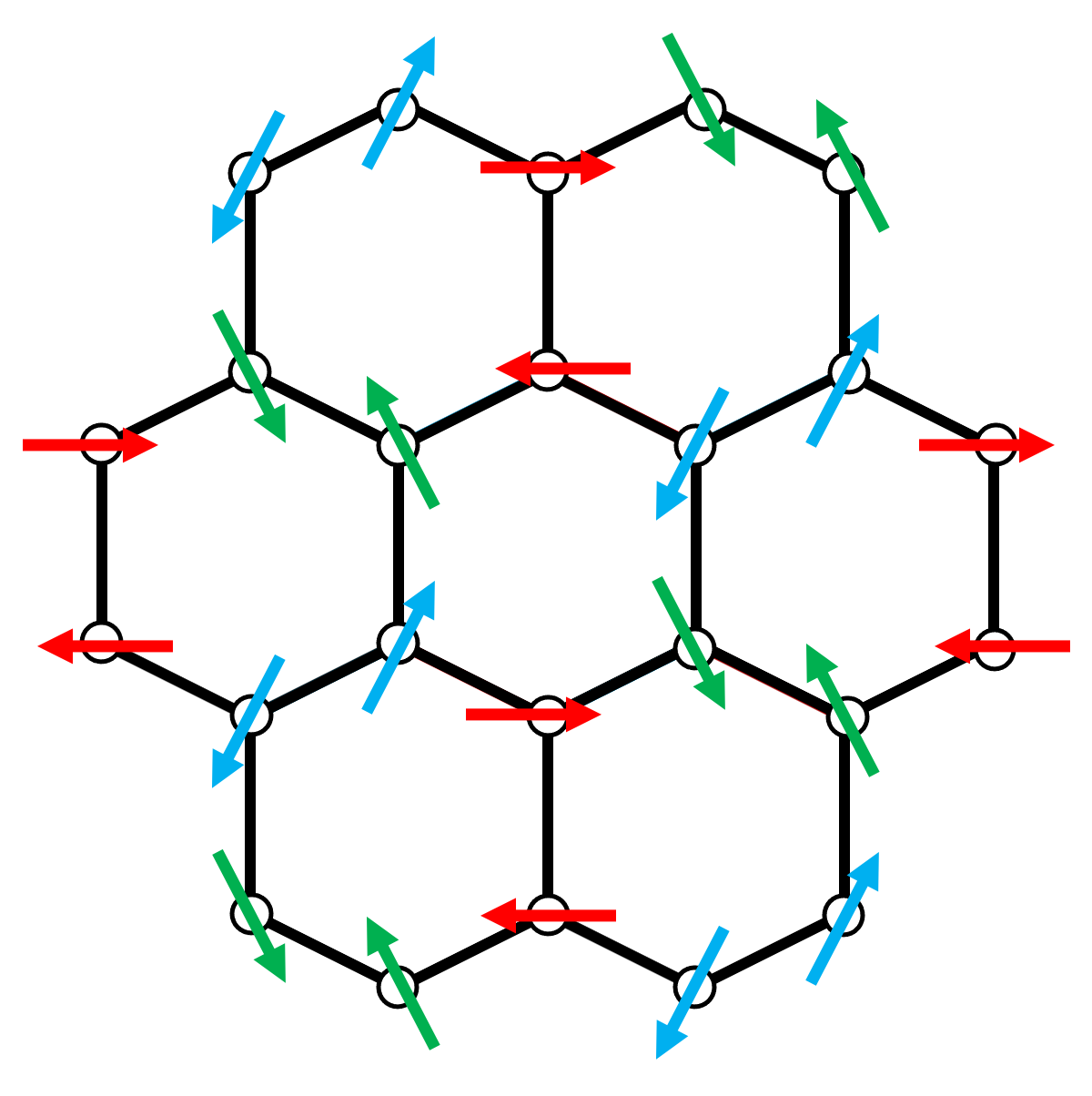}}}
    \hspace*{\fill}%

    \caption{\captiontitle{Magnetic phases supported by the 24 site cluster.} (a) Néel Antiferromagnet. (b) Zigzag antiferromagnet. (c) Stripy antiferromagnet. (d) Ferromagnet. (e) $120^{\circ}$. QSL and spiral phases not shown.} \label{fig:putativePhases}
\end{figure*}

\begin{figure*}[h]
    \hspace*{\fill}%
    \subfloat[][]{\conditionalGraphics{\includegraphics[scale=0.5]{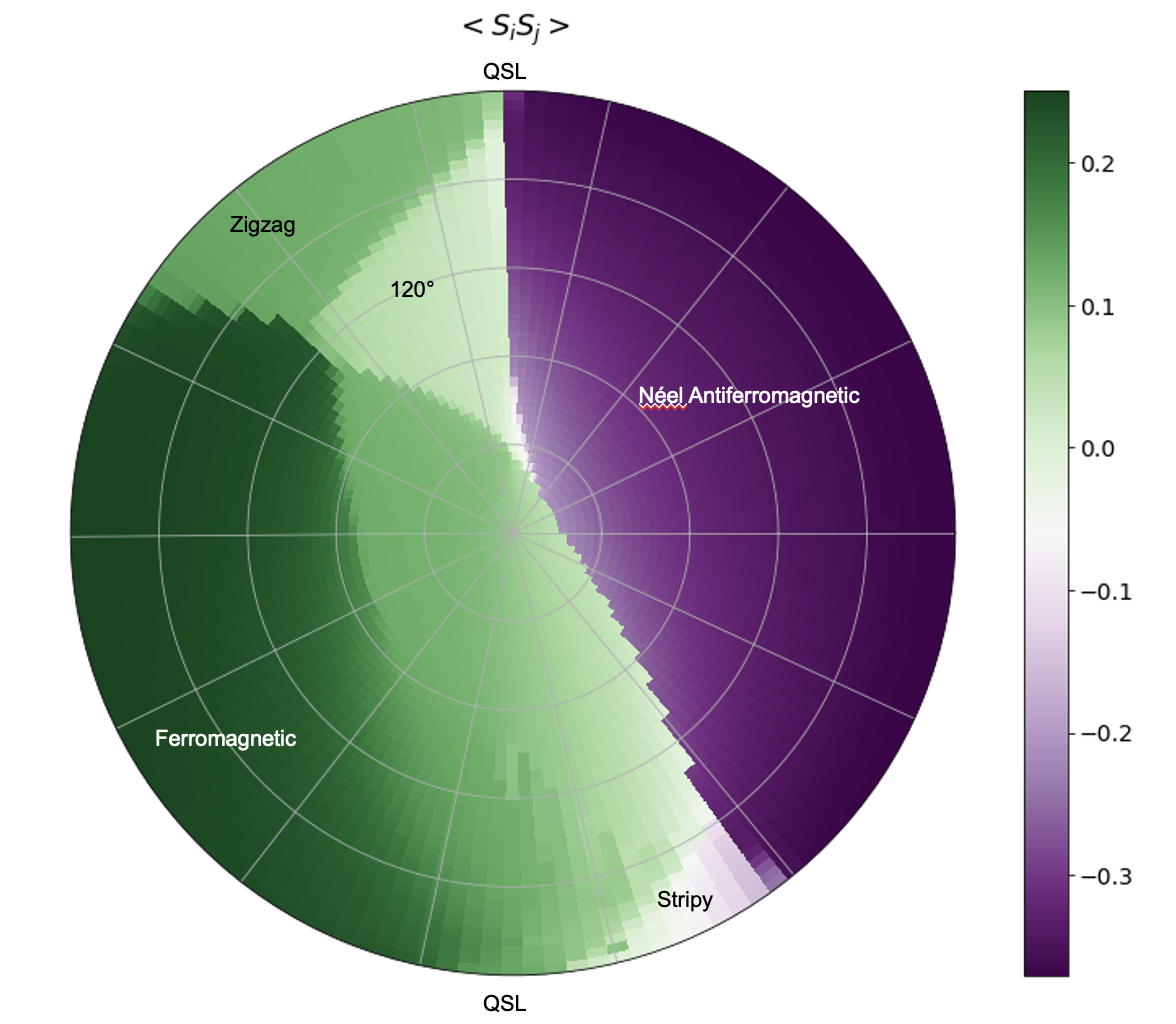}}}
    \hspace*{\fill}
    \\~\\
    \hspace*{\fill}%
    \subfloat[][]{\conditionalGraphics{\includegraphics[scale=0.2]{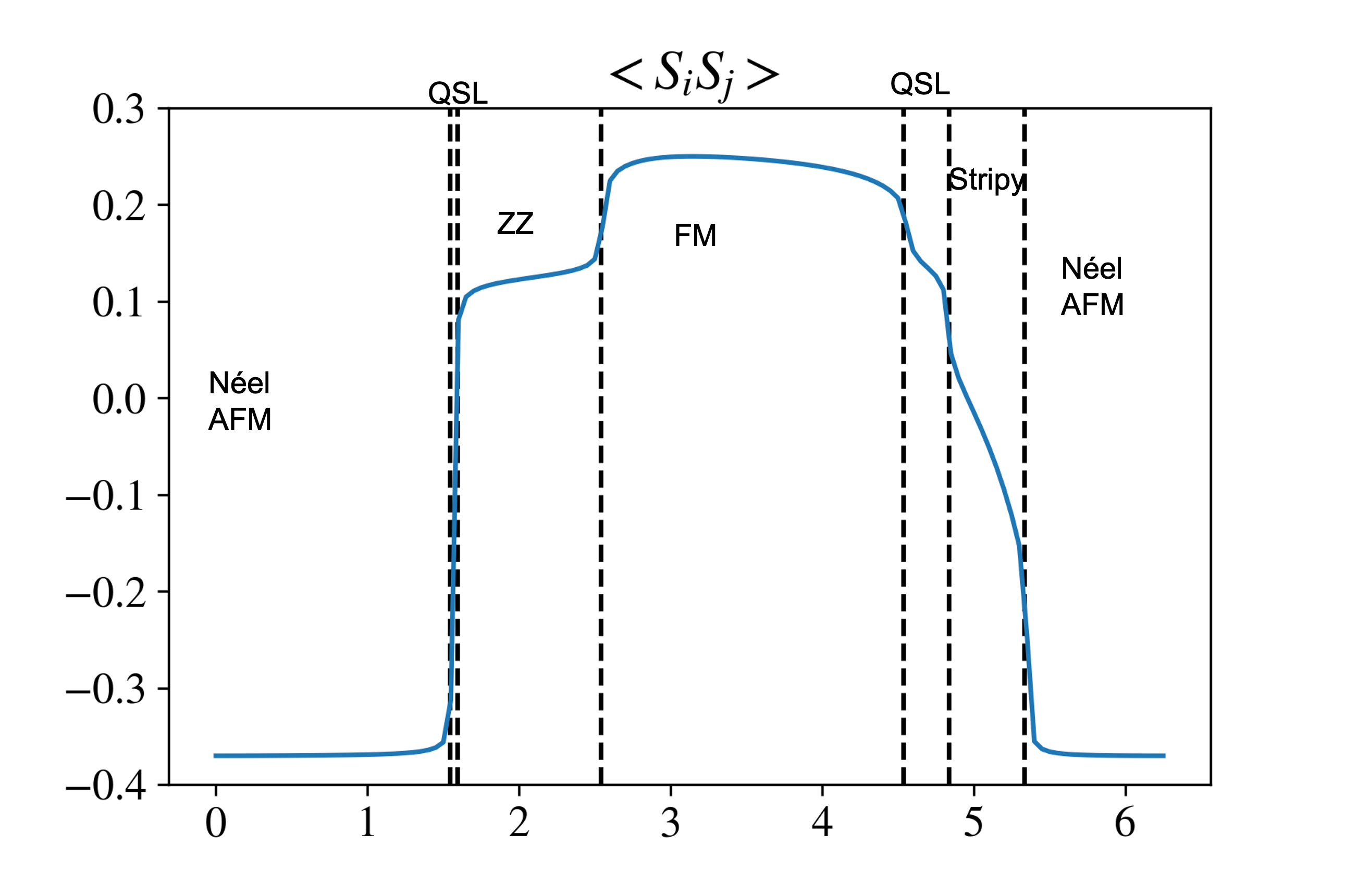}}}
    \hspace*{\fill}

    \caption{\captiontitle{Phase diagram of $JK\Gamma$ model}(a) Phase diagram (for $\Gamma \geq 0$) with the observable $\langle \mathbf{S}_i \mathbf{S}_j \rangle$. (b) Phase diagram of the model when $\theta = \pi/2$. The QSL regions are very small and nearly invisible on the full $JK\Gamma$ model but are more visible when the slice is taken.} \label{fig:phasediagramJKG}
\end{figure*}

In addition, to verify our spin model calculations, we calculate the full phase diagram of the $JK\Gamma$ model with \begin{align}
    J = \sin\theta \cos\phi \quad K = \sin\theta \sin \theta \quad \Gamma = \cos \theta
\end{align}
The 24 site cluster used in this study supports a number of magnetic phases beyond the phases merely in the Heisenberg-Kitaev model. These and the calculated phase diagram are shown in Figs. \ref{fig:putativePhases}, \ref{fig:phasediagramJKG}. We observe strong agreement between our results and previous works; for a more detailed discussion of this full phase diagram, we refer the reader to \cite{rau14, rau2014trigonal}. 

\section{Discussion of Material Parameters}
\begin{figure*}
\conditionalGraphics{\includegraphics[trim={0cm 13cm 0cm 0},clip, scale=0.5]{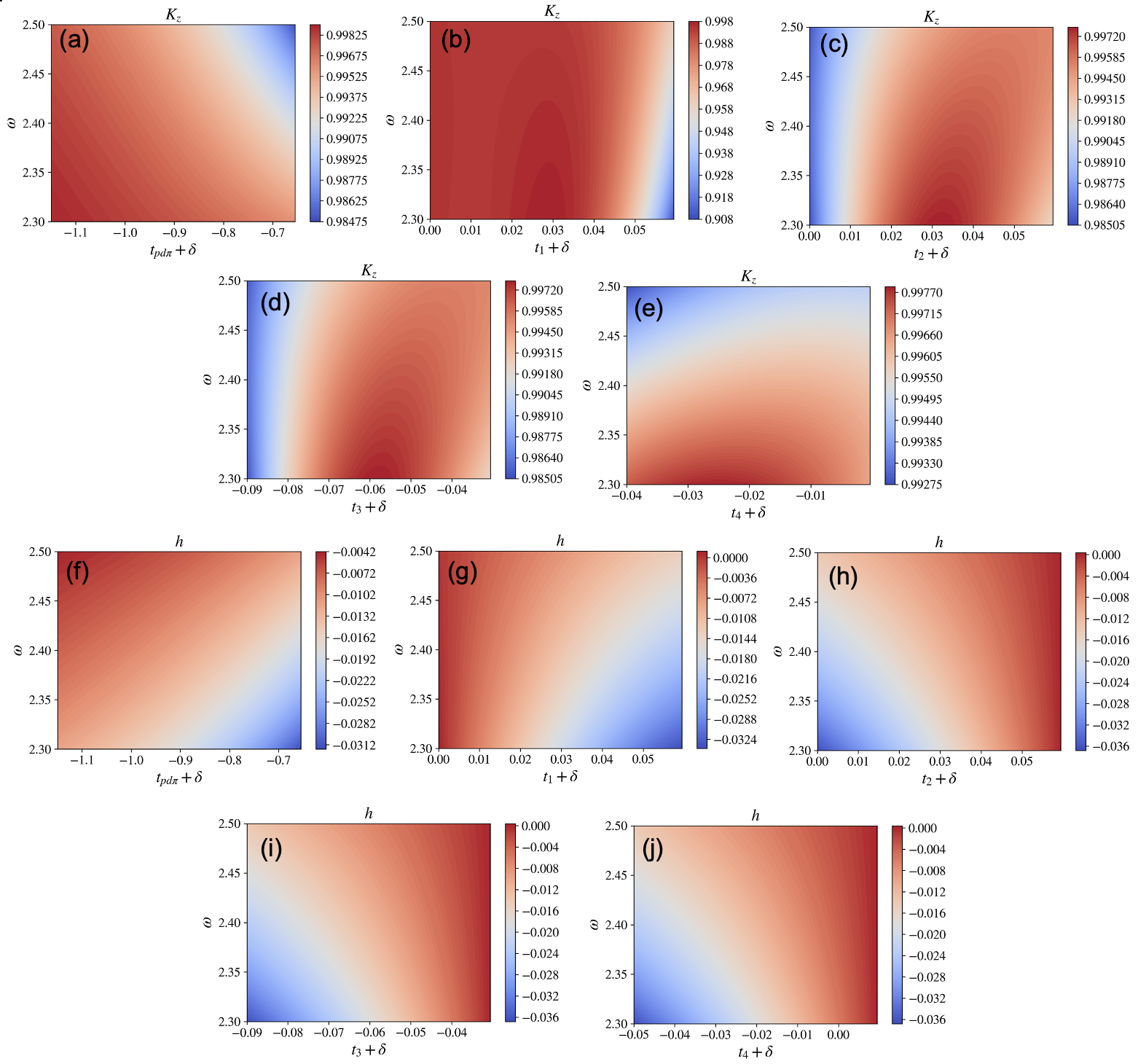}}
\caption{\captiontitle{Dependence of Kitaev exchange on electronic hopping.} For completeness, the dependence of the Floquet spin exchange parameters on variations of the electronic hopping parameters is shown due to substantial variations in reported parameters for the electronic model.} \label{fig:hoppingDependence}
\end{figure*}

There is widespread disagreement on the material parameters of both RuCl\textsubscript{3}. There have been a many studies of these materials, each using a different set of material parameters and in turn each study reports different magnetic exchange coefficients \cite{Winter_2017, winterNComm2017}. In this section, we demonstrate that the conclusions reached in the main paper do not change for different values of the material parameters. We have shown that the drive amplitude and frequency both provide handles to tune the exchange coefficients. As the material parameters change, the frequency and amplitude of the drive required to engineer a spin liquid change but the process remains. These results are shown in Fig. \ref{fig:hoppingDependence}. These results show how the magnetic exchange coefficients depend with the hopping integrals. Except for the parameter varying on the $x$-axis, all other material parameters were fixed to the RuCl\textsubscript{3} values. Although we do see variation in the magnetic parameters, around the frequency range in which the spin liquid was identified this variation is rather small. Only for significant changes in the hopping integrals do we see substantial changes in the results. This suggests that even though there exists widespread discrepancies between reported values of these numbers, our results remain robust.
\newpage

\bibliography{main}

\begin{thebibliography}{38}%
\makeatletter
\providecommand \@ifxundefined [1]{%
 \@ifx{#1\undefined}
}%
\providecommand \@ifnum [1]{%
 \ifnum #1\expandafter \@firstoftwo
 \else \expandafter \@secondoftwo
 \fi
}%
\providecommand \@ifx [1]{%
 \ifx #1\expandafter \@firstoftwo
 \else \expandafter \@secondoftwo
 \fi
}%
\providecommand \natexlab [1]{#1}%
\providecommand \enquote  [1]{``#1''}%
\providecommand \bibnamefont  [1]{#1}%
\providecommand \bibfnamefont [1]{#1}%
\providecommand \citenamefont [1]{#1}%
\providecommand \href@noop [0]{\@secondoftwo}%
\providecommand \href [0]{\begingroup \@sanitize@url \@href}%
\providecommand \@href[1]{\@@startlink{#1}\@@href}%
\providecommand \@@href[1]{\endgroup#1\@@endlink}%
\providecommand \@sanitize@url [0]{\catcode `\\12\catcode `\$12\catcode
  `\&12\catcode `\#12\catcode `\^12\catcode `\_12\catcode `\%12\relax}%
\providecommand \@@startlink[1]{}%
\providecommand \@@endlink[0]{}%
\providecommand \url  [0]{\begingroup\@sanitize@url \@url }%
\providecommand \@url [1]{\endgroup\@href {#1}{\urlprefix }}%
\providecommand \urlprefix  [0]{URL }%
\providecommand \Eprint [0]{\href }%
\providecommand \doibase [0]{https://doi.org/}%
\providecommand \selectlanguage [0]{\@gobble}%
\providecommand \bibinfo  [0]{\@secondoftwo}%
\providecommand \bibfield  [0]{\@secondoftwo}%
\providecommand \translation [1]{[#1]}%
\providecommand \BibitemOpen [0]{}%
\providecommand \bibitemStop [0]{}%
\providecommand \bibitemNoStop [0]{.\EOS\space}%
\providecommand \EOS [0]{\spacefactor3000\relax}%
\providecommand \BibitemShut  [1]{\csname bibitem#1\endcsname}%
\let\auto@bib@innerbib\@empty
\bibitem [{\citenamefont {Keimer}\ and\ \citenamefont
  {Moore}(2017)}]{keimer17}%
  \BibitemOpen
  \bibfield  {author} {\bibinfo {author} {\bibfnamefont {B.}~\bibnamefont
  {Keimer}}\ and\ \bibinfo {author} {\bibfnamefont {J.~E.}\ \bibnamefont
  {Moore}},\ }\bibfield  {title} {\bibinfo {title} {The physics of quantum
  materials},\ }\href@noop {} {\bibfield  {journal} {\bibinfo  {journal}
  {Nature Phys.}\ }\textbf {\bibinfo {volume} {13}},\ \bibinfo {pages} {1045}
  (\bibinfo {year} {2017})}\BibitemShut {NoStop}%
\bibitem [{\citenamefont {Basov}\ \emph {et~al.}(2017)\citenamefont {Basov},
  \citenamefont {Averitt},\ and\ \citenamefont {Hsieh}}]{basov17}%
  \BibitemOpen
  \bibfield  {author} {\bibinfo {author} {\bibfnamefont {D.~N.}\ \bibnamefont
  {Basov}}, \bibinfo {author} {\bibfnamefont {R.~D.}\ \bibnamefont {Averitt}},\
  and\ \bibinfo {author} {\bibfnamefont {D.}~\bibnamefont {Hsieh}},\ }\bibfield
   {title} {\bibinfo {title} {Towards properties on demand in quantum
  materials},\ }\href@noop {} {\bibfield  {journal} {\bibinfo  {journal}
  {Nature Materials}\ }\textbf {\bibinfo {volume} {16}},\ \bibinfo {pages}
  {1077} (\bibinfo {year} {2017})}\BibitemShut {NoStop}%
\bibitem [{\citenamefont {de~la Torre}\ \emph {et~al.}(2021)\citenamefont
  {de~la Torre}, \citenamefont {Kennes}, \citenamefont {Claassen},
  \citenamefont {Gerber}, \citenamefont {McIver},\ and\ \citenamefont
  {Sentef}}]{delatorre21}%
  \BibitemOpen
  \bibfield  {author} {\bibinfo {author} {\bibfnamefont {A.}~\bibnamefont
  {de~la Torre}}, \bibinfo {author} {\bibfnamefont {D.~M.}\ \bibnamefont
  {Kennes}}, \bibinfo {author} {\bibfnamefont {M.}~\bibnamefont {Claassen}},
  \bibinfo {author} {\bibfnamefont {S.}~\bibnamefont {Gerber}}, \bibinfo
  {author} {\bibfnamefont {J.~W.}\ \bibnamefont {McIver}},\ and\ \bibinfo
  {author} {\bibfnamefont {M.~A.}\ \bibnamefont {Sentef}},\ }\bibfield  {title}
  {\bibinfo {title} {Nonthermal pathways to ultrafast control in quantum
  materials},\ }\href@noop {} {\bibfield  {journal} {\bibinfo  {journal} {Rev.
  Mod. Phys.}\ }\textbf {\bibinfo {volume} {93}},\ \bibinfo {pages} {041002}
  (\bibinfo {year} {2021})}\BibitemShut {NoStop}%
\bibitem [{\citenamefont {Oka}\ and\ \citenamefont {Kitamura}(2019)}]{oka18}%
  \BibitemOpen
  \bibfield  {author} {\bibinfo {author} {\bibfnamefont {T.}~\bibnamefont
  {Oka}}\ and\ \bibinfo {author} {\bibfnamefont {S.}~\bibnamefont {Kitamura}},\
  }\bibfield  {title} {\bibinfo {title} {Floquet engineering of quantum
  materials},\ }\href@noop {} {\bibfield  {journal} {\bibinfo  {journal} {Annu.
  Rev. Condens. Mat. Phys.}\ }\textbf {\bibinfo {volume} {10}},\ \bibinfo
  {pages} {387} (\bibinfo {year} {2019})}\BibitemShut {NoStop}%
\bibitem [{\citenamefont {Wang}\ \emph {et~al.}(2013)\citenamefont {Wang},
  \citenamefont {Steinberg}, \citenamefont {Jarillo-Herrero},\ and\
  \citenamefont {Gedik}}]{wang13}%
  \BibitemOpen
  \bibfield  {author} {\bibinfo {author} {\bibfnamefont {Y.~H.}\ \bibnamefont
  {Wang}}, \bibinfo {author} {\bibfnamefont {H.}~\bibnamefont {Steinberg}},
  \bibinfo {author} {\bibfnamefont {P.}~\bibnamefont {Jarillo-Herrero}},\ and\
  \bibinfo {author} {\bibfnamefont {N.}~\bibnamefont {Gedik}},\ }\bibfield
  {title} {\bibinfo {title} {Observation of {F}loquet-{B}loch states on the
  surface of a topological insulator},\ }\href@noop {} {\bibfield  {journal}
  {\bibinfo  {journal} {Science}\ }\textbf {\bibinfo {volume} {342}},\ \bibinfo
  {pages} {453} (\bibinfo {year} {2013})}\BibitemShut {NoStop}%
\bibitem [{\citenamefont {Mahmood}\ \emph {et~al.}(2016)\citenamefont
  {Mahmood}, \citenamefont {Chan}, \citenamefont {Alpichshev}, \citenamefont
  {Gardner}, \citenamefont {Lee}, \citenamefont {Lee},\ and\ \citenamefont
  {Gedik}}]{mahmood16}%
  \BibitemOpen
  \bibfield  {author} {\bibinfo {author} {\bibfnamefont {F.}~\bibnamefont
  {Mahmood}}, \bibinfo {author} {\bibfnamefont {C.-K.}\ \bibnamefont {Chan}},
  \bibinfo {author} {\bibfnamefont {Z.}~\bibnamefont {Alpichshev}}, \bibinfo
  {author} {\bibfnamefont {D.}~\bibnamefont {Gardner}}, \bibinfo {author}
  {\bibfnamefont {Y.}~\bibnamefont {Lee}}, \bibinfo {author} {\bibfnamefont
  {P.~A.}\ \bibnamefont {Lee}},\ and\ \bibinfo {author} {\bibfnamefont
  {N.}~\bibnamefont {Gedik}},\ }\bibfield  {title} {\bibinfo {title} {Selective
  scattering between {F}loquet–{B}loch and {V}olkov states in a topological
  insulator},\ }\href@noop {} {\bibfield  {journal} {\bibinfo  {journal}
  {Nature Phys.}\ }\textbf {\bibinfo {volume} {12}},\ \bibinfo {pages} {306}
  (\bibinfo {year} {2016})}\BibitemShut {NoStop}%
\bibitem [{\citenamefont {McIver}\ \emph {et~al.}(2020)\citenamefont {McIver},
  \citenamefont {Schulte}, \citenamefont {Stein}, \citenamefont {Matsuyama},
  \citenamefont {Jotzu}, \citenamefont {Meier},\ and\ \citenamefont
  {Cavalleri}}]{mciver20}%
  \BibitemOpen
  \bibfield  {author} {\bibinfo {author} {\bibfnamefont {J.~W.}\ \bibnamefont
  {McIver}}, \bibinfo {author} {\bibfnamefont {B.}~\bibnamefont {Schulte}},
  \bibinfo {author} {\bibfnamefont {F.~U.}\ \bibnamefont {Stein}}, \bibinfo
  {author} {\bibfnamefont {T.}~\bibnamefont {Matsuyama}}, \bibinfo {author}
  {\bibfnamefont {G.}~\bibnamefont {Jotzu}}, \bibinfo {author} {\bibfnamefont
  {G.}~\bibnamefont {Meier}},\ and\ \bibinfo {author} {\bibfnamefont
  {A.}~\bibnamefont {Cavalleri}},\ }\bibfield  {title} {\bibinfo {title}
  {Light-induced anomalous hall effect in graphene},\ }\href@noop {} {\bibfield
   {journal} {\bibinfo  {journal} {Nature Phys.}\ }\textbf {\bibinfo {volume}
  {16}},\ \bibinfo {pages} {38} (\bibinfo {year} {2020})}\BibitemShut {NoStop}%
\bibitem [{\citenamefont {Bukov}\ \emph {et~al.}(2016)\citenamefont {Bukov},
  \citenamefont {Kolodrubetz},\ and\ \citenamefont {Polkovnikov}}]{bukov15b}%
  \BibitemOpen
  \bibfield  {author} {\bibinfo {author} {\bibfnamefont {M.}~\bibnamefont
  {Bukov}}, \bibinfo {author} {\bibfnamefont {M.}~\bibnamefont {Kolodrubetz}},\
  and\ \bibinfo {author} {\bibfnamefont {A.}~\bibnamefont {Polkovnikov}},\
  }\bibfield  {title} {\bibinfo {title} {Schrieffer-{W}olff transformation for
  periodically driven systems: Strongly correlated systems with artificial
  gauge fields},\ }\href@noop {} {\bibfield  {journal} {\bibinfo  {journal}
  {Phys. Rev. Lett.}\ }\textbf {\bibinfo {volume} {116}},\ \bibinfo {pages}
  {125301} (\bibinfo {year} {2016})}\BibitemShut {NoStop}%
\bibitem [{\citenamefont {Mentink}\ \emph {et~al.}(2015)\citenamefont
  {Mentink}, \citenamefont {Balzer},\ and\ \citenamefont
  {Eckstein}}]{mentink15}%
  \BibitemOpen
  \bibfield  {author} {\bibinfo {author} {\bibfnamefont {J.~H.}\ \bibnamefont
  {Mentink}}, \bibinfo {author} {\bibfnamefont {K.}~\bibnamefont {Balzer}},\
  and\ \bibinfo {author} {\bibfnamefont {M.}~\bibnamefont {Eckstein}},\
  }\bibfield  {title} {\bibinfo {title} {Ultrafast and reversible control of
  the exchange interaction in {M}ott insulators},\ }\href@noop {} {\bibfield
  {journal} {\bibinfo  {journal} {Nat. Comm.}\ }\textbf {\bibinfo {volume}
  {6}},\ \bibinfo {pages} {6708} (\bibinfo {year} {2015})}\BibitemShut
  {NoStop}%
\bibitem [{\citenamefont {Claassen}\ \emph {et~al.}(2017)\citenamefont
  {Claassen}, \citenamefont {Jiang}, \citenamefont {Moritz},\ and\
  \citenamefont {Devereaux}}]{claassen17}%
  \BibitemOpen
  \bibfield  {author} {\bibinfo {author} {\bibfnamefont {M.}~\bibnamefont
  {Claassen}}, \bibinfo {author} {\bibfnamefont {H.-C.}\ \bibnamefont {Jiang}},
  \bibinfo {author} {\bibfnamefont {B.}~\bibnamefont {Moritz}},\ and\ \bibinfo
  {author} {\bibfnamefont {T.~P.}\ \bibnamefont {Devereaux}},\ }\bibfield
  {title} {\bibinfo {title} {Dynamical time-reversal symmetry breaking and
  photo-induced chiral spin liquids in frustrated {M}ott insulators},\
  }\href@noop {} {\bibfield  {journal} {\bibinfo  {journal} {Nature Comm.}\
  }\textbf {\bibinfo {volume} {8}},\ \bibinfo {pages} {1192} (\bibinfo {year}
  {2017})}\BibitemShut {NoStop}%
\bibitem [{\citenamefont {Liu}\ \emph {et~al.}(2018)\citenamefont {Liu},
  \citenamefont {Hejazi},\ and\ \citenamefont {Balents}}]{liu18}%
  \BibitemOpen
  \bibfield  {author} {\bibinfo {author} {\bibfnamefont {J.}~\bibnamefont
  {Liu}}, \bibinfo {author} {\bibfnamefont {K.}~\bibnamefont {Hejazi}},\ and\
  \bibinfo {author} {\bibfnamefont {L.}~\bibnamefont {Balents}},\ }\bibfield
  {title} {\bibinfo {title} {Floquet engineering of multi-orbital mott
  insulators: applications to orthorhombic titanates},\ }\href@noop {}
  {\bibfield  {journal} {\bibinfo  {journal} {Phys. Rev. Lett.}\ }\textbf
  {\bibinfo {volume} {121}},\ \bibinfo {pages} {107201} (\bibinfo {year}
  {2018})}\BibitemShut {NoStop}%
\bibitem [{\citenamefont {Hejazi}\ \emph {et~al.}(019R)\citenamefont {Hejazi},
  \citenamefont {Liu},\ and\ \citenamefont {Balents}}]{hejazi18}%
  \BibitemOpen
  \bibfield  {author} {\bibinfo {author} {\bibfnamefont {K.}~\bibnamefont
  {Hejazi}}, \bibinfo {author} {\bibfnamefont {J.}~\bibnamefont {Liu}},\ and\
  \bibinfo {author} {\bibfnamefont {L.}~\bibnamefont {Balents}},\ }\bibfield
  {title} {\bibinfo {title} {Floquet spin and spin-orbital hamiltonians and
  doublon-holon generations in periodically driven mott insulators},\
  }\href@noop {} {\bibfield  {journal} {\bibinfo  {journal} {Phys. Rev. B}\
  }\textbf {\bibinfo {volume} {99}},\ \bibinfo {pages} {205111} (\bibinfo
  {year} {2019(R)})}\BibitemShut {NoStop}%
\bibitem [{\citenamefont {Chaudhary}\ \emph {et~al.}(2019)\citenamefont
  {Chaudhary}, \citenamefont {Hsieh},\ and\ \citenamefont
  {Refael}}]{chaudhary19}%
  \BibitemOpen
  \bibfield  {author} {\bibinfo {author} {\bibfnamefont {S.}~\bibnamefont
  {Chaudhary}}, \bibinfo {author} {\bibfnamefont {D.}~\bibnamefont {Hsieh}},\
  and\ \bibinfo {author} {\bibfnamefont {G.}~\bibnamefont {Refael}},\
  }\bibfield  {title} {\bibinfo {title} {Orbital floquet engineering of
  exchange interactions in magnetic materials},\ }\href
  {https://doi.org/10.1103/PhysRevB.100.220403} {\bibfield  {journal} {\bibinfo
   {journal} {Phys. Rev. B}\ }\textbf {\bibinfo {volume} {100}},\ \bibinfo
  {pages} {220403(R)} (\bibinfo {year} {2019})}\BibitemShut {NoStop}%
\bibitem [{\citenamefont {Arakawa}\ and\ \citenamefont
  {Yonemitsu}(2021)}]{arakawa21}%
  \BibitemOpen
  \bibfield  {author} {\bibinfo {author} {\bibfnamefont {N.}~\bibnamefont
  {Arakawa}}\ and\ \bibinfo {author} {\bibfnamefont {K.}~\bibnamefont
  {Yonemitsu}},\ }\bibfield  {title} {\bibinfo {title} {Floquet engineering of
  mott insulators with strong spin-orbit coupling},\ }\href
  {https://doi.org/10.1103/PhysRevB.103.L100408} {\bibfield  {journal}
  {\bibinfo  {journal} {Phys. Rev. B}\ }\textbf {\bibinfo {volume} {103}},\
  \bibinfo {pages} {L100408} (\bibinfo {year} {2021})}\BibitemShut {NoStop}%
\bibitem [{\citenamefont {Ron}\ \emph {et~al.}(2020)\citenamefont {Ron},
  \citenamefont {Chaudhary}, \citenamefont {Zhang}, \citenamefont {Ning},
  \citenamefont {Zoghlin}, \citenamefont {Wilson}, \citenamefont {Averitt},
  \citenamefont {Refael},\ and\ \citenamefont
  {Hsieh}}]{PhysRevLett.125.197203}%
  \BibitemOpen
  \bibfield  {author} {\bibinfo {author} {\bibfnamefont {A.}~\bibnamefont
  {Ron}}, \bibinfo {author} {\bibfnamefont {S.}~\bibnamefont {Chaudhary}},
  \bibinfo {author} {\bibfnamefont {G.}~\bibnamefont {Zhang}}, \bibinfo
  {author} {\bibfnamefont {H.}~\bibnamefont {Ning}}, \bibinfo {author}
  {\bibfnamefont {E.}~\bibnamefont {Zoghlin}}, \bibinfo {author} {\bibfnamefont
  {S.~D.}\ \bibnamefont {Wilson}}, \bibinfo {author} {\bibfnamefont {R.~D.}\
  \bibnamefont {Averitt}}, \bibinfo {author} {\bibfnamefont {G.}~\bibnamefont
  {Refael}},\ and\ \bibinfo {author} {\bibfnamefont {D.}~\bibnamefont
  {Hsieh}},\ }\bibfield  {title} {\bibinfo {title} {Ultrafast enhancement of
  ferromagnetic spin exchange induced by ligand-to-metal charge transfer},\
  }\href@noop {} {\bibfield  {journal} {\bibinfo  {journal} {Phys. Rev. Lett.}\
  }\textbf {\bibinfo {volume} {125}},\ \bibinfo {pages} {197203} (\bibinfo
  {year} {2020})}\BibitemShut {NoStop}%
\bibitem [{\citenamefont {Chaudhary}\ \emph {et~al.}(2020)\citenamefont
  {Chaudhary}, \citenamefont {Ron}, \citenamefont {Hsieh},\ and\ \citenamefont
  {Refael}}]{chaudhary2020controlling}%
  \BibitemOpen
  \bibfield  {author} {\bibinfo {author} {\bibfnamefont {S.}~\bibnamefont
  {Chaudhary}}, \bibinfo {author} {\bibfnamefont {A.}~\bibnamefont {Ron}},
  \bibinfo {author} {\bibfnamefont {D.}~\bibnamefont {Hsieh}},\ and\ \bibinfo
  {author} {\bibfnamefont {G.}~\bibnamefont {Refael}},\ }\bibfield  {title}
  {\bibinfo {title} {Controlling ligand-mediated exchange interactions in
  periodically driven magnetic materials},\ }\href@noop {} {\bibfield
  {journal} {\bibinfo  {journal} {arXiv:2009.00813}\ } (\bibinfo {year}
  {2020})}\BibitemShut {NoStop}%
\bibitem [{\citenamefont {Jackeli}\ and\ \citenamefont
  {Khaliullin}(2009)}]{jackeli09}%
  \BibitemOpen
  \bibfield  {author} {\bibinfo {author} {\bibfnamefont {G.}~\bibnamefont
  {Jackeli}}\ and\ \bibinfo {author} {\bibfnamefont {G.}~\bibnamefont
  {Khaliullin}},\ }\bibfield  {title} {\bibinfo {title} {Mott insulators in the
  strong spin-orbit coupling limit: From {H}eisenberg to a quantum compass and
  {K}itaev models},\ }\href@noop {} {\bibfield  {journal} {\bibinfo  {journal}
  {Phys. Rev. Lett.}\ }\textbf {\bibinfo {volume} {102}},\ \bibinfo {pages}
  {017205} (\bibinfo {year} {2009})}\BibitemShut {NoStop}%
\bibitem [{\citenamefont {Chaloupka}\ \emph {et~al.}(2013)\citenamefont
  {Chaloupka}, \citenamefont {Jackeli},\ and\ \citenamefont
  {Khaliullin}}]{chaloupka12}%
  \BibitemOpen
  \bibfield  {author} {\bibinfo {author} {\bibfnamefont {J.}~\bibnamefont
  {Chaloupka}}, \bibinfo {author} {\bibfnamefont {G.}~\bibnamefont {Jackeli}},\
  and\ \bibinfo {author} {\bibfnamefont {G.}~\bibnamefont {Khaliullin}},\
  }\bibfield  {title} {\bibinfo {title} {Zigzag magnetic order in the iridium
  oxide {N}a\textsubscript{2}{I}r{O}\textsubscript{3}},\ }\href@noop {}
  {\bibfield  {journal} {\bibinfo  {journal} {Phys. Rev. Lett.}\ }\textbf
  {\bibinfo {volume} {110}},\ \bibinfo {pages} {097204} (\bibinfo {year}
  {2013})}\BibitemShut {NoStop}%
\bibitem [{\citenamefont {Rau}\ \emph {et~al.}(2014)\citenamefont {Rau},
  \citenamefont {Lee},\ and\ \citenamefont {Kee}}]{rau14}%
  \BibitemOpen
  \bibfield  {author} {\bibinfo {author} {\bibfnamefont {J.~G.}\ \bibnamefont
  {Rau}}, \bibinfo {author} {\bibfnamefont {E.~K.-H.}\ \bibnamefont {Lee}},\
  and\ \bibinfo {author} {\bibfnamefont {H.-Y.}\ \bibnamefont {Kee}},\
  }\bibfield  {title} {\bibinfo {title} {Generic spin model for the honeycomb
  iridates beyond the {K}itaev limit},\ }\href@noop {} {\bibfield  {journal}
  {\bibinfo  {journal} {Phys. Rev. Lett.}\ }\textbf {\bibinfo {volume} {112}},\
  \bibinfo {pages} {077204} (\bibinfo {year} {2014})}\BibitemShut {NoStop}%
\bibitem [{\citenamefont {Winter}\ \emph {et~al.}(2016)\citenamefont {Winter},
  \citenamefont {Li}, \citenamefont {Jeschke},\ and\ \citenamefont
  {Valent\'{\i}}}]{PhysRevB.93.214431}%
  \BibitemOpen
  \bibfield  {author} {\bibinfo {author} {\bibfnamefont {S.~M.}\ \bibnamefont
  {Winter}}, \bibinfo {author} {\bibfnamefont {Y.}~\bibnamefont {Li}}, \bibinfo
  {author} {\bibfnamefont {H.~O.}\ \bibnamefont {Jeschke}},\ and\ \bibinfo
  {author} {\bibfnamefont {R.}~\bibnamefont {Valent\'{\i}}},\ }\bibfield
  {title} {\bibinfo {title} {Challenges in design of {K}itaev materials:
  Magnetic interactions from competing energy scales},\ }\href@noop {}
  {\bibfield  {journal} {\bibinfo  {journal} {Phys. Rev. B}\ }\textbf {\bibinfo
  {volume} {93}},\ \bibinfo {pages} {214431} (\bibinfo {year}
  {2016})}\BibitemShut {NoStop}%
\bibitem [{\citenamefont {Winter}\ \emph
  {et~al.}(2017{\natexlab{a}})\citenamefont {Winter}, \citenamefont {Tsirlin},
  \citenamefont {Daghofer}, \citenamefont {van~den Brink}, \citenamefont
  {Singh}, \citenamefont {Gegenwart},\ and\ \citenamefont
  {Valent{\'{\i}}}}]{Winter_2017}%
  \BibitemOpen
  \bibfield  {author} {\bibinfo {author} {\bibfnamefont {S.~M.}\ \bibnamefont
  {Winter}}, \bibinfo {author} {\bibfnamefont {A.~A.}\ \bibnamefont {Tsirlin}},
  \bibinfo {author} {\bibfnamefont {M.}~\bibnamefont {Daghofer}}, \bibinfo
  {author} {\bibfnamefont {J.}~\bibnamefont {van~den Brink}}, \bibinfo {author}
  {\bibfnamefont {Y.}~\bibnamefont {Singh}}, \bibinfo {author} {\bibfnamefont
  {P.}~\bibnamefont {Gegenwart}},\ and\ \bibinfo {author} {\bibfnamefont
  {R.}~\bibnamefont {Valent{\'{\i}}}},\ }\bibfield  {title} {\bibinfo {title}
  {Models and materials for generalized {K}itaev magnetism},\ }\href@noop {}
  {\bibfield  {journal} {\bibinfo  {journal} {Journal of Physics: Condensed
  Matter}\ }\textbf {\bibinfo {volume} {29}},\ \bibinfo {pages} {493002}
  (\bibinfo {year} {2017}{\natexlab{a}})}\BibitemShut {NoStop}%
\bibitem [{\citenamefont {Gotfryd}\ \emph {et~al.}(2017)\citenamefont
  {Gotfryd}, \citenamefont {Rusna\ifmmode~\check{c}\else \v{c}\fi{}ko},
  \citenamefont {Wohlfeld}, \citenamefont {Jackeli}, \citenamefont
  {Chaloupka},\ and\ \citenamefont {Ole\ifmmode~\acute{s}\else
  \'{s}\fi{}}}]{PhysRevB.95.024426}%
  \BibitemOpen
  \bibfield  {author} {\bibinfo {author} {\bibfnamefont {D.}~\bibnamefont
  {Gotfryd}}, \bibinfo {author} {\bibfnamefont {J.}~\bibnamefont
  {Rusna\ifmmode~\check{c}\else \v{c}\fi{}ko}}, \bibinfo {author}
  {\bibfnamefont {K.}~\bibnamefont {Wohlfeld}}, \bibinfo {author}
  {\bibfnamefont {G.}~\bibnamefont {Jackeli}}, \bibinfo {author} {\bibfnamefont
  {J.~c.~v.}\ \bibnamefont {Chaloupka}},\ and\ \bibinfo {author} {\bibfnamefont
  {A.~M.}\ \bibnamefont {Ole\ifmmode~\acute{s}\else \'{s}\fi{}}},\ }\bibfield
  {title} {\bibinfo {title} {Phase diagram and spin correlations of the
  {K}itaev-{H}eisenberg model: Importance of quantum effects},\ }\href@noop {}
  {\bibfield  {journal} {\bibinfo  {journal} {Phys. Rev. B}\ }\textbf {\bibinfo
  {volume} {95}},\ \bibinfo {pages} {024426} (\bibinfo {year}
  {2017})}\BibitemShut {NoStop}%
\bibitem [{\citenamefont {Sears}\ \emph {et~al.}(2015)\citenamefont {Sears},
  \citenamefont {Songvilay}, \citenamefont {Plumb}, \citenamefont {Clancy},
  \citenamefont {Qiu}, \citenamefont {Zhao}, \citenamefont {Parshall},\ and\
  \citenamefont {Kim}}]{PhysRevB.91.144420}%
  \BibitemOpen
  \bibfield  {author} {\bibinfo {author} {\bibfnamefont {J.~A.}\ \bibnamefont
  {Sears}}, \bibinfo {author} {\bibfnamefont {M.}~\bibnamefont {Songvilay}},
  \bibinfo {author} {\bibfnamefont {K.~W.}\ \bibnamefont {Plumb}}, \bibinfo
  {author} {\bibfnamefont {J.~P.}\ \bibnamefont {Clancy}}, \bibinfo {author}
  {\bibfnamefont {Y.}~\bibnamefont {Qiu}}, \bibinfo {author} {\bibfnamefont
  {Y.}~\bibnamefont {Zhao}}, \bibinfo {author} {\bibfnamefont {D.}~\bibnamefont
  {Parshall}},\ and\ \bibinfo {author} {\bibfnamefont {Y.-J.}\ \bibnamefont
  {Kim}},\ }\bibfield  {title} {\bibinfo {title} {Magnetic order in
  $\alpha$-{RuCl}\textsubscript{3}: A honeycomb-lattice quantum magnet with
  strong spin-orbit coupling},\ }\href@noop {} {\bibfield  {journal} {\bibinfo
  {journal} {Phys. Rev. B}\ }\textbf {\bibinfo {volume} {91}},\ \bibinfo
  {pages} {144420} (\bibinfo {year} {2015})}\BibitemShut {NoStop}%
\bibitem [{Note1()}]{Note1}%
  \BibitemOpen
  \bibinfo {note} {For simplicity, we impose $C_3$ symmetry on the hopping
  matrix elements, however a generalization to account for distortions of the
  $C2/m$ space group is straightforward.}\BibitemShut {Stop}%
\bibitem [{Note2()}]{Note2}%
  \BibitemOpen
  \bibinfo {note} {E.g. a dimensionless field strength of $\protect \bar {A} =
  0.25$ at a frequency $\hbar \Omega = 2.5$ eV corresponds to an electric field
  amplitude of $\protect \mathcal {E} \approx 1.8 \times 10^7$ V $\cdot $
  cm$^{-1}$}\BibitemShut {NoStop}%
\bibitem [{Note3()}]{Note3}%
  \BibitemOpen
  \bibinfo {note} {The explicit Fourier series expansion is provided in the
  Supplementary Material for completeness.}\BibitemShut {Stop}%
\bibitem [{sup()}]{supp}%
  \BibitemOpen
  \href@noop {} {\bibinfo {title} {See supplementary info}}\BibitemShut
  {NoStop}%
\bibitem [{Note4()}]{Note4}%
  \BibitemOpen
  \bibinfo {note} {Throughout this study the values of these constants used
  were adapted from recent \protect \textit {ab initio} and photoemission
  studies: $U = 3.0, J_H = 0.45, U' = U-2J_H, \lambda = 0.15, \Delta = 5, t_1 =
  0.03, t_2 = 0.03, t_3 = -0.06, t_4 = -0.02$ and $t_{pd} = -0.9$ \cite
  {PhysRevB.93.155143, sinn2016}.}\BibitemShut {Stop}%
\bibitem [{Note5()}]{Note5}%
  \BibitemOpen
  \bibinfo {note} {Normalized values for $J, \Gamma , \Gamma '$ are provided in
  the supplementary material for completeness.}\BibitemShut {Stop}%
\bibitem [{\citenamefont {Wang}\ \emph {et~al.}(2017)\citenamefont {Wang},
  \citenamefont {Claassen}, \citenamefont {Moritz},\ and\ \citenamefont
  {Devereaux}}]{wang17}%
  \BibitemOpen
  \bibfield  {author} {\bibinfo {author} {\bibfnamefont {Y.}~\bibnamefont
  {Wang}}, \bibinfo {author} {\bibfnamefont {M.}~\bibnamefont {Claassen}},
  \bibinfo {author} {\bibfnamefont {B.~J.}\ \bibnamefont {Moritz}},\ and\
  \bibinfo {author} {\bibfnamefont {T.~P.}\ \bibnamefont {Devereaux}},\
  }\bibfield  {title} {\bibinfo {title} {Producing coherent excitations in
  pumped mott antiferromagnetic insulators},\ }\href@noop {} {\bibfield
  {journal} {\bibinfo  {journal} {Phys. Rev. B}\ }\textbf {\bibinfo {volume}
  {96}},\ \bibinfo {pages} {235142} (\bibinfo {year} {2017})}\BibitemShut
  {NoStop}%
\bibitem [{\citenamefont {Winter}\ \emph
  {et~al.}(2017{\natexlab{b}})\citenamefont {Winter}, \citenamefont {Reidl},
  \citenamefont {Maksimov}, \citenamefont {Chernyshev}, \citenamefont
  {Honecker},\ and\ \citenamefont {Valentí}}]{winterNComm2017}%
  \BibitemOpen
  \bibfield  {author} {\bibinfo {author} {\bibfnamefont {S.~M.}\ \bibnamefont
  {Winter}}, \bibinfo {author} {\bibfnamefont {K.}~\bibnamefont {Reidl}},
  \bibinfo {author} {\bibfnamefont {P.~A.}\ \bibnamefont {Maksimov}}, \bibinfo
  {author} {\bibfnamefont {A.~L.}\ \bibnamefont {Chernyshev}}, \bibinfo
  {author} {\bibfnamefont {A.}~\bibnamefont {Honecker}},\ and\ \bibinfo
  {author} {\bibfnamefont {R.}~\bibnamefont {Valentí}},\ }\bibfield  {title}
  {\bibinfo {title} {Breakdown of magnons in a strongly spin-orbital coupled
  magnet},\ }\href@noop {} {\bibfield  {journal} {\bibinfo  {journal} {Nature
  Comm.}\ }\textbf {\bibinfo {volume} {8}},\ \bibinfo {pages} {1152} (\bibinfo
  {year} {2017}{\natexlab{b}})}\BibitemShut {NoStop}%
\bibitem [{\citenamefont {Gordon}\ \emph {et~al.}(2019)\citenamefont {Gordon},
  \citenamefont {Catuneanu}, \citenamefont {S{\o}rensen},\ and\ \citenamefont
  {Kee}}]{Gordon2019}%
  \BibitemOpen
  \bibfield  {author} {\bibinfo {author} {\bibfnamefont {J.}~\bibnamefont
  {Gordon}}, \bibinfo {author} {\bibfnamefont {A.}~\bibnamefont {Catuneanu}},
  \bibinfo {author} {\bibfnamefont {E.}~\bibnamefont {S{\o}rensen}},\ and\
  \bibinfo {author} {\bibfnamefont {H.~Y.}\ \bibnamefont {Kee}},\ }\bibfield
  {title} {\bibinfo {title} {Theory of the field-revealed kitaev spin liquid},\
  }\href@noop {} {\bibfield  {journal} {\bibinfo  {journal} {Nature
  Communications}\ }\textbf {\bibinfo {volume} {10}},\ \bibinfo {pages} {2470}
  (\bibinfo {year} {2019})}\BibitemShut {NoStop}%
\bibitem [{\citenamefont {Takikawa}\ and\ \citenamefont
  {Fujimoto}(2019)}]{PhysRevB.99.224409}%
  \BibitemOpen
  \bibfield  {author} {\bibinfo {author} {\bibfnamefont {D.}~\bibnamefont
  {Takikawa}}\ and\ \bibinfo {author} {\bibfnamefont {S.}~\bibnamefont
  {Fujimoto}},\ }\bibfield  {title} {\bibinfo {title} {Impact of off-diagonal
  exchange interactions on the kitaev spin-liquid state of
  $\alpha$-{RuCl}\textsubscript{3}},\ }\href@noop {} {\bibfield  {journal}
  {\bibinfo  {journal} {Phys. Rev. B}\ }\textbf {\bibinfo {volume} {99}},\
  \bibinfo {pages} {224409} (\bibinfo {year} {2019})}\BibitemShut {NoStop}%
\bibitem [{\citenamefont {Kawakami}\ \emph {et~al.}(2018)\citenamefont
  {Kawakami}, \citenamefont {Itoh}, \citenamefont {Yonemitsu},\ and\
  \citenamefont {Iwai}}]{kawakami}%
  \BibitemOpen
  \bibfield  {author} {\bibinfo {author} {\bibfnamefont {Y.}~\bibnamefont
  {Kawakami}}, \bibinfo {author} {\bibfnamefont {H.}~\bibnamefont {Itoh}},
  \bibinfo {author} {\bibfnamefont {K.}~\bibnamefont {Yonemitsu}},\ and\
  \bibinfo {author} {\bibfnamefont {S.}~\bibnamefont {Iwai}},\ }\bibfield
  {title} {\bibinfo {title} {Strong light-field effects driven by nearly
  single-cycle 7 fs light-field in correlated organic conductors},\ }\href@noop
  {} {\bibfield  {journal} {\bibinfo  {journal} {J. Phys. B: At. Mol. Opt.
  Phys.}\ }\textbf {\bibinfo {volume} {51}},\ \bibinfo {pages} {174005}
  (\bibinfo {year} {2018})}\BibitemShut {NoStop}%
\bibitem [{\citenamefont {Kennes}\ \emph {et~al.}(2018)\citenamefont {Kennes},
  \citenamefont {de~la Torre}, \citenamefont {Ron}, \citenamefont {Hsieh},\
  and\ \citenamefont {Millis}}]{kennes18}%
  \BibitemOpen
  \bibfield  {author} {\bibinfo {author} {\bibfnamefont {D.~M.}\ \bibnamefont
  {Kennes}}, \bibinfo {author} {\bibfnamefont {A.}~\bibnamefont {de~la Torre}},
  \bibinfo {author} {\bibfnamefont {A.}~\bibnamefont {Ron}}, \bibinfo {author}
  {\bibfnamefont {D.}~\bibnamefont {Hsieh}},\ and\ \bibinfo {author}
  {\bibfnamefont {A.~J.}\ \bibnamefont {Millis}},\ }\bibfield  {title}
  {\bibinfo {title} {Floquet engineering in quantum chains},\ }\href@noop {}
  {\bibfield  {journal} {\bibinfo  {journal} {Phys. Rev. Lett.}\ }\textbf
  {\bibinfo {volume} {120}},\ \bibinfo {pages} {127601} (\bibinfo {year}
  {2018})}\BibitemShut {NoStop}%
\bibitem [{\citenamefont {Kim}\ and\ \citenamefont
  {Kee}(2016)}]{PhysRevB.93.155143}%
  \BibitemOpen
  \bibfield  {author} {\bibinfo {author} {\bibfnamefont {H.-S.}\ \bibnamefont
  {Kim}}\ and\ \bibinfo {author} {\bibfnamefont {H.-Y.}\ \bibnamefont {Kee}},\
  }\bibfield  {title} {\bibinfo {title} {Crystal structure and magnetism in
  $\alpha$-{RuCl}\textsubscript{3}: An ab initio study},\ }\href@noop {}
  {\bibfield  {journal} {\bibinfo  {journal} {Phys. Rev. B}\ }\textbf {\bibinfo
  {volume} {93}},\ \bibinfo {pages} {155143} (\bibinfo {year}
  {2016})}\BibitemShut {NoStop}%
\bibitem [{\citenamefont {Sears}\ \emph {et~al.}(2020)\citenamefont {Sears},
  \citenamefont {Chern}, \citenamefont {Kim}, \citenamefont {Bereciartua},
  \citenamefont {Francoual}, \citenamefont {Kim},\ and\ \citenamefont
  {Kim}}]{sears20}%
  \BibitemOpen
  \bibfield  {author} {\bibinfo {author} {\bibfnamefont {J.~A.}\ \bibnamefont
  {Sears}}, \bibinfo {author} {\bibfnamefont {L.~E.}\ \bibnamefont {Chern}},
  \bibinfo {author} {\bibfnamefont {S.}~\bibnamefont {Kim}}, \bibinfo {author}
  {\bibfnamefont {P.~J.}\ \bibnamefont {Bereciartua}}, \bibinfo {author}
  {\bibfnamefont {S.}~\bibnamefont {Francoual}}, \bibinfo {author}
  {\bibfnamefont {Y.~B.}\ \bibnamefont {Kim}},\ and\ \bibinfo {author}
  {\bibfnamefont {Y.~J.}\ \bibnamefont {Kim}},\ }\bibfield  {title} {\bibinfo
  {title} {Ferromagnetic {K}itaev interaction and the origin of large magnetic
  anisotropy in $\alpha$-{RuCl}$_3$},\ }\href@noop {} {\bibfield  {journal}
  {\bibinfo  {journal} {Nature Phys.}\ }\textbf {\bibinfo {volume} {16}},\
  \bibinfo {pages} {837} (\bibinfo {year} {2020})}\BibitemShut {NoStop}%
\bibitem [{\citenamefont {Sinn}\ \emph {et~al.}(2016)\citenamefont {Sinn},
  \citenamefont {Kim}, \citenamefont {Kim}, \citenamefont {Lee}, \citenamefont
  {Won}, \citenamefont {Oh}, \citenamefont {Han}, \citenamefont {Chang},
  \citenamefont {Hur}, \citenamefont {Sato}, \citenamefont {Park},
  \citenamefont {Kim}, \citenamefont {Kim},\ and\ \citenamefont
  {Noh}}]{sinn2016}%
  \BibitemOpen
  \bibfield  {author} {\bibinfo {author} {\bibfnamefont {S.}~\bibnamefont
  {Sinn}}, \bibinfo {author} {\bibfnamefont {C.~H.}\ \bibnamefont {Kim}},
  \bibinfo {author} {\bibfnamefont {B.~H.}\ \bibnamefont {Kim}}, \bibinfo
  {author} {\bibfnamefont {K.~D.}\ \bibnamefont {Lee}}, \bibinfo {author}
  {\bibfnamefont {C.~J.}\ \bibnamefont {Won}}, \bibinfo {author} {\bibfnamefont
  {J.~S.}\ \bibnamefont {Oh}}, \bibinfo {author} {\bibfnamefont
  {M.}~\bibnamefont {Han}}, \bibinfo {author} {\bibfnamefont {Y.~J.}\
  \bibnamefont {Chang}}, \bibinfo {author} {\bibfnamefont {N.}~\bibnamefont
  {Hur}}, \bibinfo {author} {\bibfnamefont {H.}~\bibnamefont {Sato}}, \bibinfo
  {author} {\bibfnamefont {B.}~\bibnamefont {Park}}, \bibinfo {author}
  {\bibfnamefont {C.}~\bibnamefont {Kim}}, \bibinfo {author} {\bibfnamefont
  {H.}~\bibnamefont {Kim}},\ and\ \bibinfo {author} {\bibfnamefont {T.~W.}\
  \bibnamefont {Noh}},\ }\bibfield  {title} {\bibinfo {title} {Electronic
  structure of the {K}itaev material $\alpha$-{RuCl}\textsubscript{3} probed by
  photoemission and inverse photoemission spectroscopies},\ }\href@noop {}
  {\bibfield  {journal} {\bibinfo  {journal} {Scientific Reports}\ }\textbf
  {\bibinfo {volume} {6}},\ \bibinfo {pages} {39544} (\bibinfo {year}
  {2016})}\BibitemShut {NoStop}%
\end{thebibliography}%

\end{document}